\documentclass[12pt,a4paper]{article}
\usepackage{latexsym}
\usepackage{epsfig,amssymb,euscript, mathrsfs}
\usepackage{amsmath}
\usepackage{hyperref}

\newcommand{\R}{\mathbb{R}}
\newcommand\Z{\mathbb{Z}}

\newcommand\diff{\mathrm{d}}
\newcommand{\de}{\partial}

\newcommand{\ii}{\mathrm{i}}

\newcommand{\ex}{\mathrm{e}}

\newcommand{\ti}{\widetilde}

\newcommand{\bea}{\begin{eqnarray}}
\newcommand{\eea}{\end{eqnarray}}
\newcommand{\be}{\begin{equation}}
\newcommand{\ee}{\end{equation}}

\newcommand{\ba}{\begin{equation}\begin{array}}
\newcommand{\ea}{\end{array}\end{equation}}

\newcommand{\nn}{\nonumber}

\newcommand{\dd}{\textrm{d}}
\def \bfone{\relax{\textrm 1\kern-.35em 1}}
\newcommand{\gag}{\mathfrak{q}}
\newcommand{\rA}{\mathring A}
\newcommand{\rV}{\mathring V}
\newcommand{\sfrad}{r_3}
\newcommand{\circrad}{r_1}

\newcommand{\kacm}{\kappa^\mathrm{ACM}}
\newcommand{\fop}{\mathcal{O}_f}
\newcommand{\bop}{\mathcal{O}_b}
\newcommand{\mkap}{m}
\newcommand{\alphab}{\alpha_b}
\newcommand{\betab}{\beta_b}
\newcommand{\gammab}{\gamma_b}
\newcommand{\harm}{\mathbf{S}}

\newcommand{\timp}{Y}
\newcommand{\JR}{J_\mathrm{R}}
\newcommand{\JFZ}{J_\mathrm{FZ}}
\newcommand{\Jsusy}{J_\mathrm{susy}}
\newcommand{\deltab}{\mu}

\newcommand{\Rch}{\mathbf R}

\numberwithin{equation}{section}

\begin{document}

\begin{titlepage}

\begin{center}

\begin{flushright}
KCL-MTH-14-22
\end{flushright}

\today

\vskip 2.8 cm

{\Large \bf Comments on the Casimir energy\\[4mm]

 in supersymmetric field theories} 

\vskip 14mm

Jakob Lorenzen and  Dario Martelli

\vskip 1cm

\textit{Department of Mathematics, King's College, London, \\
The Strand, London WC2R 2LS,  United Kingdom\\}

\

\end{center}

\vskip 2cm

\begin{abstract}
\noindent  
We study  the Casimir energy of four-dimensional  supersymmetric gauge theories in the context of the rigid limit of 
new minimal supergravity.  Firstly,  revisiting the computation of the localized partition function on $S^1\times S^3$, we recover   the supersymmetric Casimir energy from its path integral definition.
Secondly, we consider the same theories in the Hamiltonian formalism on  $\R \times S^3$, focussing on the free limit and 
including a one-parameter family of background gauge fields  along $\mathbb{R}$.
We compute the vacuum expectation value  of the canonical  Hamiltonian using zeta function regularization, and show 
that this interpolates between the supersymmetric Casimir energy and the ordinary Casimir energy of a supersymmetric free field theory.  
\end{abstract}

\end{titlepage}

\pagestyle{plain}
\setcounter{page}{1}
\newcounter{bean}
\baselineskip18pt
\tableofcontents

\section{Introduction}

It is remarkable that in certain situations the  path integral of an interacting
supersymmetric field theory can be reduced to the computation of one-loop determinants, using the idea of localization \cite{Witten:1988ze}.
Employing this method, the partition function $Z$ of ${\cal N}=1$  supersymmetric gauge theories with an $R$-symmetry, defined on a Hopf surface 
$S^1\times M_3$, with periodic boundary conditions for the fermions on $S^1$, has been computed  in  \cite{Assel:2014paa}.  This is   proportional 
\cite{Closset:2013vra} to the supersymmetric index   \cite{Romelsberger:2005eg}
\bea
{\cal I} (\beta) & = & \mathrm{Tr} \, (-1)^F \ex^{-\beta H_\mathrm{susy}}~,
\eea
where $F$ is the fermion number and  $H_\mathrm{susy}$ is the Hamiltonian generating translations on $S^1$, and commuting with at least one supercharge $Q$,  
which exists on the manifold $S^1\times M_3$. Here  the trace is over the Hilbert space of states on $M_3$.    
Although this index can be refined introducing additional fugacities, associated to other 
conserved charges commuting with $Q$,  below we  consider  the simplest case with only one fugacity $\beta$,    proportional to the radius  of the $S^1$. 

In analogy with the standard path integral definition of the vacuum energy of a field theory, 
from the localized partition function $Z$, one can define a quantity dubbed \emph{supersymmetric Casimir energy} \cite{Cassani:2014zwa,Assel:2014paa}   as
\bea
E_\mathrm{susy} & \equiv  & - \lim_{\beta\to \infty} \frac{\dd}{\dd \beta} \log Z  ~.
\label{firstesusy}
\eea
In \cite{Assel:2014paa} this was shown to be given by a linear combination
of the central charges  $\mathbf a$  and $\mathbf c$ of the supersymmetric theory, which in the simplest case reads
\bea
E_\mathrm{susy} & = & \frac{4}{27}(\mathbf a+3 \mathbf c) ~.
\label{secondesusy}
\eea
In two dimensions, the fact that  the Casimir energy is proportional to the unique central charge $\mathbf{c}$ has been known for long time \cite{Bloete:1986qm}, and attempts
to generalize this result to higher dimensions have been discussed in several places \cite{Cappelli:1988vw,Herzog:2013ed,Brown:1977sj}. The 
usual strategy  consists in   relating  the expectation value $\langle T_{tt}\rangle $ of the energy-momentum tensor to its  anomalous trace $\langle T^\mu_\mu \rangle$.  However, this approach 
has some limitations. First of all, the  resulting Casimir energy is \emph{ambiguous}, with the ambiguities being related to the ambiguities in the trace anomaly.  Moreover,  previous results focussed 
on the case of conformally flat geometries.\footnote{While here we also restrict ourselves to a round metric on $S^3$, the methods we use do not rely on conformal 
invariance and can be extended to the more general definition of the supersymmetric Casimir energy on a Hopf surface  \cite{Assel:2014paa}.}
Finally, 
in a generic background with non-dynamical fields, the energy-momentum tensor is not conserved, even classically.

One may be concerned that  the definition (\ref{firstesusy}) leads to an ambiguous result, due to the existence of local counterterms, that can shift arbitrarily the value of $\log Z$. 
 However,  it has been shown in  \cite{Assel:2014tba} that in the context of new minimal supergravity \cite{Sohnius:1981tp}, on $S^1\times M_3$ all the possible supersymmetric counterterms vanish, strongly suggesting that the supersymmetric Casimir energy is 
not ambiguous.    Thus, for supersymmetric theories, we  regard (\ref{secondesusy}) as a natural  generalization of the results \cite{Bloete:1986qm} in two dimensions.
Notice that this is an exact result,  thus valid for any value of the coupling constants.  For example, it is insensitive to the superpotential of the theory. 

The linear combination  of   central charges $\mathbf a$, $\mathbf c$
in  (\ref{secondesusy}) had previously  appeared in  \cite{Kim:2012ava}, in the context of studies of the supersymmetric index. Recent  papers exploring 
relations of the central charges $\mathbf a$, $\mathbf c$  with the supersymmetric index include   \cite{Buican:2014qla,DiPietro:2014bca, Ardehali:2014esa}.

In this note we investigate further the supersymmetric Casimir energy.  
In particular, we show that it coincides with  the vacuum expectation value (vev) of the Hamiltonian $H_\mathrm{susy}$, 
 appearing  in the definition of the supersymmetric index 
 \bea
\langle H_\mathrm{susy}  \rangle & =  &  E_\mathrm{susy}  ~ ,
\label{HEsusy}
\eea
computed using zeta function regularization.

Since the results (\ref{secondesusy}) and  (\ref{HEsusy})  are also  valid  at weak (or zero) coupling,  
one may compare these with the Casimir energy   in free supersymmetric 
field theories, computed using zeta function techniques.   
In particular,  for a free theory of $N_v$ vector multiplets and $N_\chi$  chiral multiplets, this reads \cite{birrell,Cappelli:1988vw}
\bea
E_\mathrm{free} \ \, =\ \, \langle H_\mathrm{free} \rangle & = & \frac{1}{192}  \left(21  N_v +5 N_\chi   \right) ~.
\label{efree}
\eea
This is the free field value of the Casimir energy  of  $N_\chi$ conformally coupled complex scalar fields, $N_\chi+N_v$ Weyl spinors, 
and $N_v$  Abelian gauge fields, and does \emph{not} agree with (\ref{HEsusy}) \cite{Cassani:2014zwa,Assel:2014paa}.  See  e.g. \cite{Marino:2011nm} for a concise derivation. 
 Working in the framework of rigid  new minimal supergravity, we will consider  the  canonical Hamiltonian $H$ of  
 a multi-parameter family of supersymmetric theories defined on the round $\mathbb{R}\times S^3$ or  
 its compactification to $S^1 \times S^3$. We will show  that this interpolates continuously 
 between $H_\mathrm{susy}$ and $H_\mathrm{free}$, thus resolving the apparent tension between    (\ref{HEsusy}) and  (\ref{efree}).  
 Generically, the Hamiltonian is not a BPS quantity,  
  and we find that its vev, computed using zeta function regularization, 
 cannot be expressed as a linear combination of the anomaly coefficients  $\mathbf a$, $\mathbf c$.

The rest of this note is organized as follows. In section \ref{susy:section} we introduce the background geometries and the field theories.  In section 
\ref{sec:path integral} we reconsider the computation of the supersymmetric Casimir energy from its path integral definition. Section 
\ref{ham:section} contains the main results of this note. We examine the field theories in the canonical formalism and derive (\ref{HEsusy}).  
Our conclusions  are presented in section \ref{concl:section}. Three appendices are included. In appendix \ref{App: Harmonics} we collect
details of the relevant spherical harmonics on the three-sphere. 
Appendix  \ref{hurwapp} contains 
the definition and some useful properties of the Hurwitz zeta function.
In appendix \ref{emten:appendix} we write expressions for the energy-momentum tensor and other useful formulas.

\section{Supersymmetric field theories}
\label{susy:section}

In this section we present the background geometry, that we view as a solution to the rigid limit of new supergravity and then introduce the relevant 
supersymmetric Lagrangians. We follow verbatim the notation of \cite{Assel:2014paa}, to which we refer for more details.

\subsection{Background geometry}
\label{sec:background}

We begin with the background in Euclidean signature, discussing the differences in Lorentzian signature later. 
We consider a background comprising the following metric
\bea \label{metric}
\dd s^2 (S^1\times S^3) & = &  \circrad^2 \dd \tau^2 + \dd s^2 (S^3)\nonumber\\
 & = &  \circrad^2 \dd \tau^2 + \frac{\sfrad^2}{4} \left( \dd \theta^2 + \sin\theta^2 \dd \varphi^2 + (\dd \varsigma + \cos\theta \dd \varphi)^2\right)~,
 \label{genrmetric}
\eea
where $\tau $ is a coordinate on $S^1$ and $\theta, \varphi, \varsigma$ with 
$0 \le \theta < \pi ~, \varphi \sim \varphi + 2 \pi ~,  \varsigma \sim \varsigma + 4 \pi $ are coordinates on a round 
three-sphere.\footnote{For  $\circrad=1$ and $\sfrad=2$ this metric and the other background fields can be obtained specializing the background discussed in
 appendix  C of  \cite{Assel:2014paa}  to $v=1$,  $b_1=-b_2=1/2$. Below we will set $\circrad=1$ and $\sfrad=2$, but these can be easily restored 
 by dimensional analysis.} We note the Ricci scalar of this metric is given  by  $R  = 6/\sfrad^2$,  which is the same as the Ricci scalar of the 3d metric. 
We introduce the following orthonormal frame\footnote{Note that this
frame is \emph{different} from the frame used in \cite{Assel:2014paa}.} 
\bea
e^1 &=& \frac{\sfrad}{2} \left( \cos \varsigma \dd \theta + \sin \theta \sin \varsigma \dd \varphi  \right)  ~,\nn\\
e^2 &=&\frac{\sfrad}{2} \left(  - \sin \varsigma \dd \theta + \sin \theta \cos \varsigma \dd \varphi \right)  ~,\nn\\
e^3 &=& \frac{\sfrad}{2} \left( \dd \varsigma + \cos \theta \dd \varphi  \right) ~,\nn\\
e^4 & =& \circrad \dd \tau  ~,\label{leftframe}
\eea
which corresponds to a \emph{left-invariant} frame $\{e^1,e^2,e^3\}$ on $S^3$. We will consider a class of backgrounds admitting 
a solution to the new minimal supersymmetry  equation 
\be
\left(\nabla_\mu - \ii A_\mu + \ii V_\mu + \ii V^\nu \sigma_{\mu\nu} \right) \zeta \ =  \ 0 ~ ,
\label{rigidsusy}
\ee
with the metric (\ref{genrmetric}).  Let us set  $\circrad=1$ and $\sfrad=2$ below.  In our coordinates  the supersymmetric complex Killing vector $K$ reads
\be
K = \frac12 \left(  \de_\varsigma - \ii \de_\tau \right)~,
\ee
and the dual one-form is 
\be
K \ = \  \frac{1}{2}(e^3-\ii e^4)~. \label{Koneform}
\ee
We define the following ``reference'' values of the background fields 
\bea
\rA & = & \frac{3}{4} e^3    +  \frac{\ii}{2}(\gag-\frac{1}{2}) e^4  ~ ,\qquad \qquad \rV  \ \, = \ \, \frac{1}{2} e^3 ~,
\eea
where we have included a constant $\gag$, which can be obtained performing a (large) gauge transformation $A \to A +  \tfrac{\ii}{2}\gag \dd \tau$ starting from the gauge choice adopted in 
\cite{Assel:2014paa}. Although in Euclidean signature this yields an ill-defined spinor, this is not true in  Lorentzian signature and later in the paper
this parameter will play a role.  Assuming that $U = \kappa K$,  where $\kappa$ is a \emph{constant}, we can write the background fields as 
\bea
A & = & \rA  +\frac{3}{2}\kappa K ~,\qquad \qquad V \ \, = \ \, \rV + \kappa K~.
\eea
 We also note the value of  the combination
\bea
A^\mathrm{cs} & = & A - \frac{3}{2}V \, =\, \rA - \frac{3}{2}\rV\, =\,   \frac{\ii}{2}(\gag-\frac{1}{2}) e^4 ~,
\eea
that is independent of $\kappa$. For generic values of $\kappa$, the solution $\zeta$ to (\ref{rigidsusy}) reads 
\be
\zeta \ = \ \frac{1  }{\sqrt 2 }\ex^{-\frac12 \gag \tau} \left( \begin{array}{c} 0 \\ 1\end{array} \right) ~ ,  \label{simpleKS}
\ee
where the normalization is chosen such that for $\gag=0$ one has $|\zeta|^2 =1/2$ as in \cite{Assel:2014paa}. Indeed, because $\tau $ is a periodic coordinate, 
this does not make sense unless\footnote{Although an appropriately  quantized pure imaginary value of $\gag$ would be allowed in (\ref{simpleKS}), for generic $R$-charges 
we must have  $\gag=0$ for the correct periodicity of the matter fields \cite{Assel:2014tba}.} $\gag=0$.

For generic values of $\kappa$ this background preserves only a $SU(2)\times U(1)$ subgroup of the isometry group $SO(4)$ of the round three-sphere.  
In the following, we will be interested in two special choices of $\kappa$.  In particular, the choice 
$\kappa  = \kacm  \equiv  -1/3$
corresponds (for $\gag=0$) to the values  in   \cite{Assel:2014paa}, namely 
\bea
A^\mathrm{ACM} & = & \frac12 e^3 ~,\qquad \qquad V^\mathrm{ACM} \ = \ \frac13 \left( e^3 + \frac{\ii }{2}  e^4 \right)~,
\eea
where notice that $A^\mathrm{ACM}$ is \emph{real}.   Another distinguished  choice is 
$\kappa  = \kappa^\mathrm{st}  \equiv  - 1$,
where the superscript stands for ``standard'', giving  
\bea
A^\mathrm{st} & = &  \frac{\ii}{2} (1+\gag) e^4 ~,\qquad \qquad V^\mathrm{st}  \ = \ \frac{ \ii}{2}e^4 ~. \label{avst}
\eea
For this  choice, the full $SO(4)$ symmetry  of the three-sphere is restored and (\ref{rigidsusy}) has the more general solution
 \be
\zeta \ = \ \ex^{-\frac12 \gag \tau} \zeta_0 ~ ,
\ee
with $\zeta_0$ any constant spinor \cite{Sen:1985dc,Romelsberger:2005eg}.

Notice that, in addition to $\gag=0$ \cite{Assel:2014paa}, there are two other special values of the parameter $\gag$. 
Namely, for $\gag =-1$ the background field  $A^\mathrm{st}$ in (\ref{avst})  vanishes, while for  $\gag =1/2$ we have  $A^\mathrm{cs}=0$. 
The significance of these three values will become clearer in later sections.

\subsection{Lagrangians}
\label{sec:Lagrangians}

We consider an ${\cal N}=1$ supersymmetric field theory with a vector multiplet transforming in the adjoint representation of a gauge group $G$, 
and  a  chiral multiplet transforming in a representation ${\cal R}$, with  Lagrangians given  in 
section 2.2 of \cite{Assel:2014paa}, evaluated in the background described in the previous section. These Lagrangians then depend on the two constant 
parameters $\gag$ and $\kappa$, as well as on the $R$-charges $r_I$ of the scalar fields in the chiral multiplet, that below will be simply denoted  $r$.  
We will restrict attention to Lagrangians expanded up to  \emph{quadratic} order in the fluctuations around a 
configuration where all fields vanish.

Adopting the notation of  \cite{Assel:2014paa}, we will therefore consider the following Lagrangian of a chiral multiplet
\bea
 {\cal L}^\mathrm{chiral}  &  = & \left( \delta_\zeta V_1 + \delta_\zeta V_2 + \epsilon \, \delta_\zeta V_U \right) |_\mathrm{quadratic} \nonumber\\[2mm]
 &= & D_{\mu}\ti \phi D^{\mu} \phi   + \left(V^{\mu} + (\epsilon-1)U^\mu \right)\left( \ii  D_{\mu}\ti \phi\,\phi - \ii \ti \phi D_{\mu}\phi \right)+ \frac{r}{4} \left( R + 6 V_{\mu}V^{\mu} \right) \ti\phi \phi     \nonumber\\
& & +           \ii \ti \psi \,\ti\sigma^{\mu}D_{\mu}\psi  + \left( \frac{1}{2} V^\mu + (1-\epsilon)U^\mu \right) \ti\psi\, \ti \sigma_{\mu} \psi  ~,
  \label{lagdef}
\eea
where $D_\mu =\nabla_\mu - \ii q_R A_\mu$ and $q_R$ denotes  the $R$-charges of the fields \cite{Assel:2014paa}.
This is therefore the Lagrangian of $|{\cal R}| \equiv N_\chi$  free chiral multiplets, each with $R$-charge $r$, 
where we will denote as $N_\chi$ the dimension of the representation ${\cal R}$.

The three terms in the first line of (\ref{lagdef}) are separately $\delta_\zeta$-exact, and the parameter $\epsilon$
allows us to continuously interpolate between the localizing Lagrangian in   \cite{Assel:2014paa}, obtained for $\epsilon=0$, and the usual chiral multiplet Lagrangian \cite{Sohnius:1981tp}, obtained for $\epsilon=1$.
Notice that at quadratic order the term  $\delta_\zeta V_3$  \cite{Assel:2014paa} vanishes. 
Inserting the values of the background fields, and writing 
\be
 {\cal L}^\mathrm{chiral}  (\gag,\kappa,\epsilon,r )   \ = \  {\cal L}^\mathrm{chiral}_\mathrm{bos}  (\gag,\kappa,\epsilon,r )  +  {\cal L}^\mathrm{chiral}_\mathrm{fer}  (\gag,\kappa,\epsilon,r ) \, , 
\ee
the bosonic part of the Lagrangian reads
\bea
{\cal L}^\mathrm{chiral}_\mathrm{bos}  (\gag,\kappa,\epsilon,r ) & = &  -\ti \phi \de_\tau^2 \phi   + \left[  \frac{r }{2} \left( 1   - 2 \gag \right)  + \kappa \left(\frac{3}{2}r-\epsilon\right) \right]\ti \phi  \de_\tau \phi
-\ti \phi \nabla^i \nabla_i \phi \nonumber\\
& & + \ii \left[ \frac{3}{2}r-1  + \kappa   \left( \frac{3}{2}  r- \epsilon \right)  \right] \ti \phi  \nabla_\varsigma \phi \nonumber\\
 & & +    \frac{r}{2} (1+\gag ) \left[ \frac{r}{2} (2-\gag ) +  \kappa \left(  \frac{3}{2}  r- \epsilon  \right)\right] \ti \phi \phi ~,\label{completebosonic}
\eea
where $\nabla_i$,  is the covariant derivative on the three-sphere, and we have omitted a total derivative.  The fermionic part of the Lagrangian reads
\bea
\mathcal L^\mathrm{chiral}_\mathrm{fer} (\gag, \kappa,\epsilon,r)  & = &   \ti \psi    \de_\tau \psi -   \ii \ti \psi   \gamma^a \de_a \psi   -   \frac{1}{2}  \left[ \frac{3}{2} r  - 1  +  \kappa \left(\frac{3}{2}r-\epsilon \right)\right] \ti \psi    \gamma_\varsigma  \psi \nonumber\\
&& -  \frac{1}{2} \left[ \frac{1}{2}(r-1) (1- 2 \gag) + \frac32 +   \kappa \left(\frac{3}{2}  r- \epsilon  \right) \right] \ti \psi \psi  ~, \label{completefermionic}
\eea
where $a=1,2,3$ are  frame indices on the three-sphere and $\gamma^a$ denote the Pauli matrices, generating the  three-dimensional Clifford algebra.
This expression is frame-dependent, and we used the left-invariant frame (\ref{leftframe}), which is  useful for applying the angular momentum formalism.  In particular, we used the identity 
\be
\ii \ti \sigma^\mu \nabla_\mu \psi \, =\,   \de_\tau \psi   - \ii\gamma^a \nabla_a \psi \ = \    \de_\tau \psi  - \ii \gamma^a \de_a \psi -\frac{3}{4}  \psi ~.
\ee
Notice that  the Lagrangians  in  \cite{Romelsberger:2005eg,Romelsberger:2007ec} correspond to the values $\epsilon=1$,  $\kappa =-1$, and $\gag=1/2$. Notice also that  for   $r=2/3$ and $\epsilon=1$ the total chiral multiplet Lagrangian does not depend on $\kappa$.

Let us introduce a compact notation, 
writing the Lagrangians above in terms of differential operators.
Denoting $\ell_a$ the Killing vectors dual to the left-invariant frame $e^a$, 
and defining the ``orbital'' angular momentum operators as $L_a = \tfrac{\ii}{2}\ell_a$, one finds these satisfy the $SU(2)$ commutation relations
 \be
\left[L_a ,L_b\right] = \ii \epsilon_{abc} L_c ~,
\ee
and we have\footnote{Recall that here we have set $\sfrad=2$. In general, the three-dimensional Laplace operator is  $\sfrad^2 \nabla^i\nabla_i \, =\, \sum_a  (\ell_a )^2$.}
 $- \nabla^i\nabla_i  =  \vec{L}^2$ and  $\nabla_\varsigma =  - \ii L_3$.
Similarly, we identify the  Pauli matrices with the spin operator as $S^a = \frac{1}{2} \gamma^a$, satisfying the same $SU(2)$ algebra.  Thus the Lagrangians  can be written as 
\bea
{\cal L}^\mathrm{chiral}_\mathrm{bos}  & = &  \ti \phi\, \widetilde {\cal O}_b\, \phi \ = \ \ti \phi \left( -\de_\tau^2 +  2\deltab \de_\tau + \bop \right) \phi~,\nn\label{grandbosonic}\\
{\cal L}^\mathrm{chiral}_\mathrm{fer} & = &  \ti \psi \, \widetilde {\cal O}_f \, \psi \ = \ \ti \psi \left( \de_\tau + \fop \right) \psi  ~,\label{grandfermionic}
\eea
where 
\bea
\bop & =  & 2 \alphab \vec L^2 + 2 \betab L_3 + \gammab~,\label{bloppy} \nn\\
\fop & =  & 2 \alpha_f \vec{L}\cdot \vec{S} + 2\beta_f S_3 + \gamma_f~,
\label{floppy}
\eea
with the constants taking the values   $\alpha_b  = \tfrac{1}{2}$,   
\bea
  \beta_b & = & - \frac{1}{2}   + \frac{3}{4}r + \frac{\kappa }{2}  \left( \frac{3}{2}  r- \epsilon \right)  ~, \nonumber\\
\gamma_b & = & \frac{r}{2} (1+\gag ) \left[ \frac{r}{2} (2-\gag ) +   \kappa \left(  \frac{3}{2}  r- \epsilon  \right)\right]  ~,\nonumber\\
\deltab & = & \frac{1}{2} \left[  \frac{r }{2} \left( 1   - 2 \gag \right)  +  \kappa \left(\frac{3}{2}r-\epsilon\right) \right]~,
\label{listbosonic}
 \eea
and  $\alpha_f  = -1$, $\beta_f=-\beta_b$, 
\bea
\gamma_f &=&  - \left[ \frac{1}{4}(r-1) (1- 2 \gag) + \frac34 +\frac{  \kappa}{2} \left(\frac{3}{2}  r- \epsilon  \right) \right] ~, \label{ABG}
\eea 
 respectively.

For the vector multiplet   the quadratic Lagrangian is 
\bea  
{\cal L}^{\rm vector}  & = &  \mathrm{Tr} \Big[  \frac {1}{4}  {\cal F}_{\mu\nu}  {\cal F}^{\mu\nu}  
 + \frac{\ii}{2} \lambda\, \sigma^{\mu} D_{\mu}^\mathrm{cs} \ti\lambda +  \frac{\ii}{2} \ti\lambda\, \ti\sigma^{\mu} D^\mathrm{cs}_{\mu}\lambda   \Big]_\mathrm{quadratic}~,
  \label{Vvector}
  \eea
where $D^\mathrm{cs}_\mu= \nabla_\mu - \ii q_R A_\mu^\mathrm{cs}$. ${\cal F}$ is the linearized field strength of the gauge field ${\cal A}$
  and  $\lambda$ is the gaugino, both transforming in the adjoint representation of the gauge group $G$.  This is therefore the Lagrangian of $|G|\equiv N_v$ free vector 
  multiplets, where we will denote $N_v$ the dimension of the gauge group $G$.

The fermionic part of this Lagrangian can be put in the same form as the fermionic part 
of the chiral multiplet Lagrangian, namely 
\bea
{\cal L}^\mathrm{vector}_\mathrm{fer} & = &  \ti \lambda \, \widetilde {\cal O}_f^\mathrm{vec} \, \lambda \ = \ \ti \lambda \left( \de_\tau + \fop^\mathrm{vec} \right) \lambda  ~,
\eea 
where 
\bea
\fop^\mathrm{vec} & =  & 2 \alpha_v \vec{L}\cdot \vec{S} + 2\beta_v S_3 + \gamma_v~,  \label{Ofvector}
\eea
with $\alpha_v=-1$, $\beta_v=0$, $\gamma_v= \frac{\gag}{2}-1$. Notice that for $\gag=1/2$, corresponding to  $A^\mathrm{cs}=0$,
this reduces to the standard massless Dirac operator on the three-sphere.

\section{Supersymmetric Casimir energy}
\label{sec:path integral}

In this section we will  recover in our set-up  the supersymmetric Casimir energy defined in  \cite{Assel:2014paa} as 
\bea
E_\mathrm{susy} & = & - \lim_{\beta\to \infty} \frac{\dd}{\dd \beta} \log Z (\beta) ~,
\label{defesusy}
\eea
where  $Z$ is the \emph{supersymmetric partition function}, namely the path integral on $S^1\times S^3$ with periodic boundary conditions for the fermions on $S^1$, computed using localization. 
Restoring the radii of $S^1$ and $S^3$, the dimensionless parameter $\beta$  in  \cite{Assel:2014paa} is given by
\be
\beta = \frac{2\pi \circrad }{\sfrad}~.
\ee
Differently from   \cite{Assel:2014paa}, here  we will not fix the value of $\kappa$, showing that owing to the pairing of bosonic and fermionic eigenvalues in the  one-loop 
determinant, the final result will be \emph{independent} of $\kappa$. Although the computation in Euclidean signature requires to fix $\gag=0$, we will start 
presenting the explicit eigenvalues for generic values of $\gag$. We will then demonstrate  that the pairing occurs if and only if $\gag=0$.

The partition function takes the form   \cite{Assel:2014paa} 
\bea
Z  (\beta) &  = & \ex^{- {\cal F} (\beta)} \, {\cal I}(\beta)~,
\eea
where ${\cal I}(\beta)$ is the supersymmetric index, and the pre-factor ${\cal F} (\beta)= - \ii \pi \left( \Psi^{(0)}_\mathrm{chi}  + \Psi^{(0)}_\mathrm{vec} \right)$ arises from the 
regularization of one-loop determinants in the chiral multiplets and vector multiplets, respectively  \cite{Assel:2014paa}
(see also \cite{Closset:2013sxa}).  Since the index  ${\cal I}(\beta)$ does not contribute to (\ref{defesusy}), in order to 
compute $E_\mathrm{susy}$ we can restrict attention to $\Psi^{(0)}_\mathrm{chi}$  and  $\Psi^{(0)}_\mathrm{vec}$, and thus effectively set the constant gauge field 
${\cal A}_0=0$ in the one-loop determinants around the localization locus in  \cite{Assel:2014paa}. In particular, as the vector multiplet Lagrangian does not depend on $\kappa$ and $\epsilon$, its contribution to $E_\mathrm{susy}$   
can be simply borrowed from  \cite{Assel:2014paa}.  For example, by setting $|b_1|=|b_2|=\circrad/\sfrad$ in eq. (4.33) of   \cite{Assel:2014paa}, one obtains
\be
\Psi^{(0)}_\mathrm{vec}   \ = \  \frac{\ii}{6}\left( \frac{\circrad}{\sfrad} - \frac{\sfrad}{\circrad}\right) N_v~.
\label{acmpsivec}
\ee

For  the chiral multiplet, we revisit the computation of the one-loop determinant (with ${\cal A}_0=0$) by working out the explicit eigenvalues for an arbitrary choice of the parameters
$\kappa$ and $\epsilon$. The eigenvalues of the operators $\bop$ and $\fop$ can be obtained with elementary methods from the theory of angular momentum in quantum mechanics
\cite{Kapustin:2009kz}. See appendix \ref{App: Harmonics} for a summary of the relevant spherical harmonics. Thus, writing 
\bea
\bop \phi & = & E^2_b \phi~,\nonumber\\
\fop \psi & = & \lambda^\pm \psi~, 
\eea
for the scalar harmonics we have 
\bea
E^2_b & =  & \frac{\alphab}{2}\ell  (\ell +2) + 2 \betab \mkap + \gammab ~,
\label{boseing}
\eea
where  $\tfrac{\ell}{2}(\tfrac{\ell}{2}+1)$   for $\ell=0,1,2,\dots$   are the eigenvalues of  $\vec{L}^2$, and
 $\mkap=-\tfrac{\ell}{2},\dots, \tfrac{\ell}{2}$,  are the eigenvalues of $L_3$. Each eigenvalue has degeneracy $(\ell +1)$, due to the $SU(2)_R$  symmetry.

We distinguish two types  of eigenvalues of $\fop$. For any $\ell = 1,2,3,...$  we have 
\bea
\lambda^\pm_{\ell m}   & = &  -\frac{\alpha_f}{2} +\gamma_f \pm \sqrt{\frac{\alpha_f^2}{4} (\ell +1)^2 + \alpha_f\beta_f (1+ 2 \mkap) + \beta_f^2}~,
\label{fermeigens}
\eea
where here the quantum number $\mkap$ takes the values 
$\mkap = -\tfrac{\ell}{2}, \dots , \tfrac{\ell}{2} - 1$.
Furthermore,  for any $\ell=0,1,2,\dots$, we have the two special eigenvalues 
\be
\lambda_\ell^\mathrm{special\pm} \, = \,  \frac{\alpha_f}{2} \ell \pm \beta_f +\gamma_f  ~. \label{special}
\ee
Again, each eigenvalue has degeneracy $(\ell +1)$, due to the $SU(2)_R$  symmetry. Expanding the fields in Kaluza-Klein modes on $S^1$ as
\bea
\phi (x) & = & \sum_{k\in \Z} \mathrm{e}^{-\ii k \tau} \phi_k (\theta,\varphi, \varsigma)
\eea
and similarly for $\psi$, we  obtain the following eigenvalues  for each mode
\bea
\widetilde  {\cal O}_b  \phi_k  & = & \left(   k^2  -2 \ii \deltab  k   + E^2_b\right)  \phi_k ~,  \nonumber\\
\widetilde {\cal O}_f \psi_k & = & \left(  -\ii k + \lambda^\pm \right) \psi_k~.
\eea
For generic values of the quantum numbers $\ell, \mkap$, we say that the eigenvalues of the operators $\widetilde  {\cal O}_b$ and  $\widetilde {\cal O}_f$
are \emph{paired}, if for all $k$ we have
\bea
\left(  -\ii k + \lambda^+ \right)  \left(  -\ii  k + \lambda^-\right) & = & - \left(  k^2   -2 \ii\deltab  k    + E^2_b\right) ~.
\label{thepair}
\eea
Inserting the values of the parameters given in (\ref{listbosonic}) and (\ref{ABG}) we find that this is satisfied if and only if $\gag=0$, and for any value of $\kappa, \epsilon, r$.
Let us then set $\gag=0$ in the rest of this section. Restoring generic values of the radius $\sfrad$ of the $S^3$,  the one-loop determinant for a fixed $k$ is 
\bea
Z^{(k)}_{\mathrm{1-loop}} & = & \frac{\det \widetilde {\cal O}_f }{\det \widetilde {\cal O}_b } \ = \  \frac{\prod_{\lambda^-} \left(  - \ii k +  \frac{2}{\sfrad} \lambda^- \right) \prod_{\lambda^+} \left(  - \ii k +  \frac{2}{\sfrad} \lambda^+ \right) }{\prod_{E_b}  \left(  k^2   - \frac{4}{\sfrad} \ii \deltab  k    + \frac{4}{\sfrad^2}E^2_b\right) }~,
\label{oneloop}
\eea
where the products are over all the bosonic and fermionic eigenvalues, including the special ones. However, using the condition (\ref{thepair}) all the paired eigenvalues
 cancel out.\footnote{Up to an irrelevant overall sign.}
For $\mkap=\ell/2$ the generic fermionic eigenvalues do not exist, thus there are \emph{unpaired} bosonic eigenvalues,  obtained setting $\mkap= \ell/2$ 
in   (\ref{boseing}), which  read 
\bea
(E_b^2)^\mathrm{unpaired} & = &  \left(\frac{\alpha_f}{2}(\ell+1)+\beta_f \right)^2 - \deltab^2\quad \qquad \ell=0,1,2,\dots~,
\eea
and remain in the denominator of (\ref{oneloop}). Therefore, taking into account the contribution of the special fermionic eigenvalues in the numerator,  and including the degeneracies, 
we obtain
\be \label{partition1}
Z_\mathrm{1-loop}^{(k)} \ = \ \prod_{n_0=1}^\infty \left(\frac{n_0+1+ \sfrad \ii  k-r}{n_0-1- \sfrad \ii k+r}\right)^{n_0}~,
\ee
where we defined $n_0=\ell+1$  and used that   $\alpha_f=-1$ and
\be
 \beta_f + \deltab \ = \ \frac{1}{2}(1-r)~.
 \ee
  Upon obvious identifications, this coincides with the one-loop 
determinant of a $d=3$, ${\cal N}=2$ chiral multiplet on the round three-sphere,  originally derived in \cite{Jafferis:2010un} and \cite{Hama:2010av}, although our
 operators $\bop$ and $\fop$ are slightly more general and interpolate between those used in these two references. In particular, the Lagrangians
 used in  \cite{Jafferis:2010un} correspond to  $\kappa = -1/3$ and $\epsilon=0$, precisely as in \cite{Assel:2014paa}, while those used in 
\cite{Hama:2010av} correspond to  $\kappa = -1$ and $\epsilon=1$. Recall that in all cases we have $\gag=0$.

Defining 
\be
z \ = \  1- r +  \frac{\sfrad \ii k}{\circrad} ~,
\ee
where  $\circrad$ can be  restored simply rescaling the coordinate $\tau\to \circrad  \tau$, 
one finds $ Z_\mathrm{1-loop}^{(k)} ( z) =   s_{b=1}(\ii z)$, where $s_b(x)$ is the double sine function \cite{Hama:2010av}.
Alternatively, (\ref{partition1}) can be written in terms of special functions by integrating the differential equation
\bea
\frac{\dd}{\dd z} \log Z_\mathrm{1-loop}^{(k)}(z)  & = & -\pi  z \cot (\pi  z)~  ,
 \label{dlogZdz}
\eea 
where the Hurwitz zeta function has been used to regularize the infinite sum  \cite{Jafferis:2010un} (see appendix \ref{hurwapp}).

In order to take the limit  $\beta \to \infty$ it is more convenient to write  (\ref{partition1}) as an infinite product over two integers, namely
\be \label{Zdoublesine}
Z_\mathrm{1-loop}^{(k)} \ = \ \prod_{n_1= 0}^\infty \prod_{n_2 = 0}^\infty \frac{n_1 +n_2+1 +z}{n_1+n_2+1-z } ~.
\ee
Regularizing  the infinite product over the Kaluza-Klein modes as in  \cite{Assel:2014paa}, one obtains $Z_\mathrm{1-loop}$ in terms of Barnes triple-gamma functions,
which eventually can be written as  
\be
Z_\mathrm{1-loop}   \  =  \ \ex^{\ii\pi  \Psi^{(0)}_\mathrm{chi}} \, \widetilde \Gamma_e \left( \frac{\ii\circrad r }{\sfrad} , \frac{\ii\circrad }{\sfrad} , \frac{\ii\circrad }{\sfrad}\right) ~ , \label{elliptic}
\ee
where  $\widetilde \Gamma_e$ is  the elliptic gamma function \cite{Assel:2014paa} and\footnote{Actually, the terms proportional to $\sfrad/\circrad$ in  (\ref{acmpsichi}) and 
(\ref{acmpsivec}) are not present if one uses a slightly different regularization, consistent with the results of \cite{DiPietro:2014bca}. However, this does not affect the terms proportional to 
$\circrad/\sfrad$, which are relevant for the computation of $E_\mathrm{susy}$. We thank B. Assel, D. Cassani, L. Di Pietro, and Z.  Komargodski for discussions on this issue. See \cite{Assel:2015nca}.} 
\be
\Psi^{(0)}_\mathrm{chi}   \ = \ \frac{\ii }{6 }  \left(\frac{2\circrad }{\sfrad}  (r-1)^3- \left(\frac{\circrad}{\sfrad} +\frac{\sfrad}{\circrad}\right)(r-1) \right)~.
\label{acmpsichi}
\ee
From this, one finds the contribution of a chiral multiplet to (\ref{defesusy}) to be
\be
 E_\mathrm{susy}^\mathrm{chiral} \ =  \ \frac{1}{12}  \left(2(r-1)^3-(r-1) \right)~.
\ee
This is exactly  the contribution of a chiral multiplet with $R$-charge $r$ to the 
total supersymmetric Casimir energy computed in  \cite{Assel:2014paa}, although we emphasize that here this  has been derived for arbitrary values of the parameters
 $\kappa$ and $\epsilon$.

Combining the contributions of the chiral multiplets and the vector multiplets  we recover the result
\bea
E_\mathrm{susy} & = & \frac{4}{27}(\mathbf a+3 \mathbf c)~,
\label{Erecover}
\eea
with the  anomaly coefficients  defined as
\bea
\mathbf a &=& \frac{3}{32} (3 \, \text{tr}\Rch^3-\text{tr}\Rch)  ~ , \qquad \mathbf c \ \, = \ \,  \frac{1}{32} (9 \, \text{tr}\Rch^3-5\, \text{tr}\Rch) ~ ,
\eea
where $\Rch$ denotes the $R$-symmetry charge, and ``tr" runs over the fermionic fields in the multiplets.

In the remainder of the paper we will show that  (\ref{Erecover}) is also equal to 
 the expectation value of the BPS Hamiltonian $H_\mathrm{susy}$,
appearing in supersymmetric index
\bea
{\cal I} (\beta ) & = & \mathrm{Tr} \, (-1)^F \ex^{-\beta H_\mathrm{susy}}~.
\eea
Therefore we now turn to the Hamiltonian formalism, working in a background with a non-compact time direction, thus with $\beta\to \infty$ from the outset.

\section{Hamiltonian formalism}
\label{ham:section}

In this section we will study the theories defined in section \ref{sec:Lagrangians} in a background $\mathbb{R}\times S^3$ in Lorentzian signature, obtained from the geometry 
 in section \ref{sec:background}  by a simple analytic continuation. In particular, we take the metric
\bea 
\dd s^2 (\R \times S^3) & = &  -  \dd t^2 + \dd s^2 (S^3)~,
\eea
where $t$ denotes the time coordinate on $\R$, and $\dd s^2 (S^3)$ is the metric on $S^3$, given in equation \eqref{genrmetric}. Below we continue to 
set $\sfrad=2$. The background fields are obtained setting $A_t = - \ii A_\tau$, $V_t = - \ii V_\tau$, and $K_t= - \ii K_\tau$, where here we must take $\kappa \in \mathbb{R}$.
Moreover, the dynamical fields obey
 $\ti \phi = \phi^\dagger$ and $\ti \psi = \psi^\dagger$ from the start. 
The $\sigma$-matrices generating  the appropriate Clifford algebra are obtained setting $\sigma^0_{\alpha \dot \alpha}=\ii \sigma^4_{\alpha \dot \alpha} = \bfone_{\alpha \dot \alpha}$ and $\ti \sigma^0_{\alpha \dot \alpha}=\ii \ti \sigma^4_{\alpha \dot \alpha} = \bfone_{\alpha \dot \alpha}$, with the remaining components unchanged, such that
\be
\sigma_a \ti \sigma_b + \sigma_b \ti \sigma_a \ = \ - 2 \eta_{ab} ~, \qquad \ti \sigma_a  \sigma_b + \ti \sigma_b \sigma_a \ = \ - 2 \eta_{ab} ~ ,
\ee
with $\eta_{ab} = \text{diag} (-1,1,1,1)$. The Lorentzian spinor $\zeta$ solving equation \eqref{rigidsusy} for generic $\kappa$ is then
\be
\zeta \ = \ \frac{\ex^{\frac12 \ii \gag t } }{\sqrt 2} \left( \begin{array}{c} 0 \\ 1 \end{array} \right) ~ ,
\ee
again with a more general solution for the special value $\kappa= \kappa^\mathrm{st} =-1$ \cite{Sen:1985dc,Romelsberger:2005eg}.

\subsection{Conserved charges}
\label{sec:Conserved Charges}

The Hamiltonian\footnote{In this section we will consider mainly the chiral multiplet Lagrangian, therefore we will drop the superscript ``chiral'' from all the quantities.} density $\mathcal H = \mathcal H_\mathrm{bos} + \mathcal H_\mathrm{fer}$, associated to the chiral multiplet 
Lagrangian (\ref{lagdef}),  is obtained as usual, by defining the canonical momenta
\be
\Pi \ =  \  \de_t \ti \phi  - \ii \deltab \ti \phi ~, \qquad \ti \Pi \  = \ \de_t \phi  + \ii \deltab \phi ~, \qquad  \pi^\alpha  \ = \ \ii \ti \psi_{\dot \alpha} \ti \sigma^{0 \, \dot \alpha \alpha}  ~ , \qquad \ \ti \pi^\alpha \ = \ 0 ~ ,
\ee
and its bosonic and fermionic parts read
\bea
\mathcal H_\mathrm{bos} &=&  \Pi \de_t \phi + \ti \Pi \de_t \ti \phi - \mathcal L^\mathrm{chiral}_\mathrm{bos}  ~, \nn  \\
\mathcal H_\mathrm{fer} & = & \pi \de_t \psi + \ti \pi \de_t \ti \psi - \mathcal L^\mathrm{chiral}_\mathrm{fer}  ~, 
\eea
respectively. In terms of the operators $\mathcal O_b$ and $\mathcal O_f$  defined in equations \eqref{bloppy} and \eqref{floppy}, 
we have
\bea
\mathcal H_\mathrm{bos} &=&   \ti \Pi \Pi  - \ii \deltab \big(   \Pi \phi- \ti \Pi \ti \phi \big) +\ti \phi\left(\mathcal O_b +\deltab^2 \right)  \phi ~, \nn \\
\mathcal H_\mathrm{fer} & = &   -\ti \psi \mathcal O_f \psi~ . \label{FermHamDen}
\eea

 The Hamiltonian is then obtained by integrating\footnote{The integral is over the spatial $S^3$ with the metric $\dd s^2(S^3)$ in \eqref{genrmetric}. We define $\dd^3 x = \dd \theta \dd \varsigma \dd \varphi$ and $g_3=\sin^2\theta$ denotes the determinant of this metric.} over the spatial $S^3$, 
\be
H   \ = \ \int \sqrt{g_3} \dd^3x \ \mathcal H ~ .
\ee
The $R$-symmetry  current $\JR^\mu$ can be derived either from the Noether procedure or as the functional derivative
 of the action with respect to $A_\mu$, namely
\bea
\JR^\mu & = & \frac{1}{\sqrt{-g}} \frac{\delta S}{\delta A_\mu} ~,
\eea
and it reads
\be
 \JR^\mu   =     \ii r \big( D^\mu \ti \phi\, \phi -  \ti \phi   D^\mu \phi \big) + 2 r \big(   V^\mu +  \kappa (\epsilon-1)   K^\mu  \big) \ti \phi \phi+  (r-1) \ti \psi \ti \sigma^\mu \psi  ~ . 
\ee
This is conserved, i.e. $\nabla_\mu \JR^\mu=0$, and the  corresponding conserved charge $R$ is obtained by contracting it with the time-like Killing vector $\de_t$, and integrating on the $S^3$, which yields
\be
R \ = \   \int \sqrt{g_3} \dd^3x \left(  \ii r   \big(\ti \phi \ti \Pi - \phi \Pi \big) ~ +  (r-1)  \ti \psi \ti \sigma^t \psi  \right).  \label{JR}
\ee
Rotational symmetry along the Killing vector $\de_\varsigma$ gives rise to a conserved current with the corresponding conserved angular momentum
\be
J_3   \ = \ - \ii \int \sqrt{g_3} \dd^3x \left(  (L_3 \phi) \, \Pi  + (L_3 \ti \phi) \, \ti \Pi  +  \ii \ti \psi \, \left( L_3 +  S_3 \right) \psi \right) ~ .  \label{JtJ3}
\ee
Finally, supersymmetry gives rise to the conserved supercurrent
\bea
\zeta^\alpha \Jsusy^\mu{\,}_\alpha & = & -   \sqrt2  \zeta \sigma^\nu \ti \sigma^\mu \psi   D_\nu\ti \phi    ~.
\eea
Using the equations of motion for the dynamical fields, after some calculations, one can verify that  
\be
\nabla_\mu (\zeta \Jsusy^\mu)  \  = \  0~.
\ee
 Note that $\nabla_\mu \zeta \neq0$, and therefore $\Jsusy^\mu$ is not conserved by itself, as is the case in the standard flat-space computation. 
 Contracting $\zeta \Jsusy^\mu$ with the time-like Killing vector  $\de_t$, we obtain the conserved supercharge
\be
Q \ = \  - \sqrt2 \int \dd^3x \sqrt{ g_3} \left(  \zeta  \psi  \, \Pi   - \ii  \ti \phi  \zeta \widehat{\mathcal O}_f   \psi  \right) ~,   \label{Q}
\ee
where we defined
\be
\widehat{\mathcal O}_f \ \equiv \  2 \widehat \alpha \vec S \cdot \vec L  +2 \widehat \beta S_3  +  \widehat \gamma  ~ ,
\ee
with
\be
\widehat \alpha \ = \ -1 ~ , \qquad \widehat \beta  \ = \    \frac34(1-r)    ~, \qquad \widehat \gamma \ = \  -   \frac{\kappa}{2} \left( \frac32 r -\epsilon \right)   -\frac34 ~ .
\ee

In summary, applying the Noether procedure to the Lagrangian  (\ref{lagdef}), 
we have derived expressions for the Hamiltonian $H$, $R$-charge $R$, angular momentum $J_3$,  and supercharge $Q$. 
These will provide the relevant operators in the quantized theory. 

Let us briefly discuss   other currents that can be considered, which however   are not conserved generically.
In particular, the  usual energy-momentum tensor,   defined as 
\be
T_{\mu \nu} \  = \   \frac{-2}{\sqrt{-g}} \frac{\delta S}{\delta g^{\mu\nu}} ~ ,
\label{def:EM tensor}
\ee
is \emph{not conserved} in the presence of non-dynamical fields.   This remains true even  if $T_{\mu\nu}$
is contracted with a vector field that generates a symmetry of the metric and the other background fields. Thus, for example, $T_{tt}$ does not define
a conserved quantity, and in particular it does not coincide with the canonical  Hamiltonian. 
Denoting the  non-dynamical vector fields as $A^I_\mu$, with $F^I= \diff A^I$,  and the associated currents as $J_I^\mu$,
in general the energy-momentum tensor (\ref{def:EM tensor}) obeys the Ward identity
\bea
\nabla^\mu T_{\mu\nu} & = & \sum_{I} \left(F^I_{\mu\nu} J_I^\mu - A^I_\nu \nabla_\mu J_I^\mu  \right) ~.
\eea
 
 In the present case, after a tedious computation,  one finds  that the energy-momentum tensor  satisfies
\bea
\nabla^\mu T_{\mu\nu} & = &    (\dd A)_{\nu \mu}  \JR^\mu  - \frac32  ( \dd V)_{\nu \mu}  \JFZ^\mu +  (\dd  K)_{\nu \mu}   J_K^\mu  \nn\\
&& +\frac32 V_\nu \nabla_\mu \JFZ^\mu -K_\nu \nabla_\mu J_K^\mu ~ ,
\label{Tnoncons}
\eea
where $\JFZ^\mu$ is the Ferrara-Zumino current
\be
 \JFZ^\mu  \ = \  -\frac{2}{3}\frac{1}{\sqrt{-g}} \frac{\delta S}{\delta V_\mu} ~ ,
\ee
and 
\be
J_K^\mu = \frac{1}{\sqrt{-g}} \frac{\delta S}{\delta K_\mu} ~ .
\ee 
Neither $\JFZ^\mu$ nor $J_K^\mu$ are conserved. Explicit expressions for $T_{\mu\nu}$, $\JFZ^\mu$, and $J_K^\mu$ are given in appendix \ref{emten:appendix}.
 Note that in this context, we must formally treat $K_\mu$ as a background field, although it was introduced in the Lagrangian as a shift of the original fields $A_\mu$ and $V_\mu$. For the usual chiral multiplet Lagrangian with $\epsilon=1$, however, one has $J_K^\mu=0$. 

For a generic Killing vector $\xi$, that is also a symmetry of the background fields, ${\cal L}_\xi A= {\cal L}_\xi V=0$,  we
can define a conserved current as
\be
\timp_\xi^\mu \ = \  \xi_\nu \left( T^{\mu\nu} + \JR^\mu A^\nu   -\frac32 \JFZ^\mu V^\nu  + J_K^\mu K^\nu   \right) ~ .
\ee
One can show that indeed $\nabla_\mu \timp_\xi^\mu=0$. In particular, for $\xi=\de_t$, one finds that the conserved charge is the Hamiltonian density
\be
\mathcal H \ =  \  - \timp_{\de_t}^t  ~ ,
\ee
up to a total derivative on the $S^3$.

\subsection{Canonical quantization}

We now expand the dynamical fields in terms of creation and annihilation operators. Let us first focus on the scalar field. 
In order for the field $\phi$ to solve its equation of motion, we expand it as
\bea
\phi (x) & = & \sum_{\ell=0}^\infty  \sum_{m,n=- \frac{\ell}{2}}^{\frac{\ell}{2}}  \left(a_{\ell  mn} u_{\ell mn}^{(+)} (x) + b^\dagger_{\ell mn}  (u_{\ell mn}^{(-)})^\ast (x)  \right)  ~,
\eea
with\footnote{Although all the eigenvalues in this paper never depend on the $SU(2)_R\subset SO(4)$ quantum number $n$, we keep track of this in the spherical harmonics and in the expansions.}
\be
 u_{\ell mn}^{(\pm)} (x) \  \equiv \ \frac{1}{4  \sqrt{\omega^\pm_{\ell m} \mp \deltab} } \ex^{-\ii  \omega_{\ell m }^{ \pm} t  } Y^{mn}_\ell (\vec x)~ , 
\ee
where $Y^{mn}_\ell (\vec x)$ are the scalar spherical harmonics on a three-sphere of unit radius (see appendix \ref{App: Harmonics} for further details), and
\be
\omega^\pm_{\ell m} \ = \ \pm \deltab + \sqrt{\frac{\alpha_b}{2} \ell (\ell+2) \pm 2 \beta_b m + \gamma_b + \deltab^2} ~ . \label{omega beta not 0}
\ee
The canonical commutation relations
\bea
[\phi (t, \vec x) , \Pi (t,\vec x\,') ] & = & \frac{\ii}{\sqrt{-g}} \delta^{(3)}(\vec x - \vec x \,')  ~ ,\nn\\
{[}\phi (t, \vec x) , \phi (t,\vec x\,')] & = & [\Pi (t, \vec x) , \Pi (t,\vec x\,')] \ =  \  0 ~ , 
\eea
with $\delta^{(3)}(\vec x - \vec x \,')= \delta(\theta- \theta') \delta (\varphi - \varphi') \delta (\varsigma - \varsigma ' )$, 
 hold by taking the oscillators to satisfy the usual
\be
[a_{\ell m n} , a^\dagger_{\ell' m' n' } ] \ = \ [b_{\ell m n} , b^\dagger_{\ell' m' n'} ] = \delta_{\ell,\ell'} \delta_{m,m'} \delta_{n,n'} ~ .
\ee
From \eqref{FermHamDen} it follows that the Hamiltonian of the scalar field reads
\bea
H_\mathrm{bos} & = & \frac{1}{2} \sum_{\ell=0}^\infty\sum_{ m,n =-\frac{\ell}{2}}^{\frac{\ell}{2}} \omega^{+}_{\ell  m }\left( a_{\ell mn } a_{\ell mn }^\dagger +a_{\ell mn }^\dagger a_{\ell mn } \right) \nn\\
& &+  \frac{1}{2} \sum_{\ell=0}^\infty\sum_{ m,n =-\frac{\ell}{2}}^{\frac{\ell}{2}} \omega^{-}_{\ell m } \left(b_{\ell mn } b_{\ell mn }^\dagger  + b^\dagger_{\ell mn } b_{\ell mn } \right)  \label{scalarHam}  ~ .
\eea
Notice that we have used the  \emph{Weyl ordering} prescription,   as this is the correct one  for comparison with the path integral approach.

For the fermion, we expand the field $\psi$ in terms of the spinor spherical harmonics $\harm^\pm_{\ell m n}$. As discussed in appendix \ref{App: Harmonics}, these are eigenspinors of the operator $\mathcal O_f$, 
\be
\mathcal O_f \harm^\pm_{\ell m n } \ =\  \lambda^\pm_{\ell m} \harm^\pm_{\ell m n }  ~ ,
\ee
with the eigenvalues $\lambda^\pm_{\ell m}$ given in equation \eqref{fermeigens}. In addition, there are the ``special'' spherical harmonics,
\bea
\mathcal O_f  \harm^\mathrm{special \pm}_{\ell n} & = & \lambda^\mathrm{special \pm}_\ell \harm^\mathrm{special\pm}_{\ell n}  ~ , 
\eea 
with $ \lambda^\mathrm{special \pm}_\ell$ given in equation \eqref{special}. We expand the field $\psi$ as
\be
\psi_\alpha \ = \   \sum_{\ell=0}^\infty \sum_{n=-\frac{\ell}{2}}^{\frac{\ell}{2}} \sum_{m=-\frac{\ell}{2}-1}^{\frac{\ell}{2}} c_{\ell m n} u_{\ell m n\, \alpha } +\sum_{\ell=1}^\infty \sum_{n=-\frac{\ell}{2}}^{\frac{\ell}{2}} \sum_{m=-\frac{\ell}{2}}^{\frac{\ell}{2}-1} d_{\ell m n}^{\dagger}  v_{\ell m n \, \alpha}  ~, \label{fermion modes}
\ee
with
\be
 u_{\ell m n \, \alpha} (x) \ = \ \frac{1}{2\sqrt2} \ex^{\ii  t \lambda^-_{\ell  m } }\harm^-_{\ell m n \, \alpha} (\vec x) ~, \qquad   \qquad  v_{\ell m n \, \alpha} (x) \ = \  \frac{1}{2\sqrt2} \ex^{\ii   t \lambda^+_{\ell m } }  \harm^+_{\ell m n\, \alpha} (\vec x)  ~ .
\ee
Here we included $\harm^\mathrm{special\pm}$ in the sums by defining
\bea
\harm^-_{\ell,\frac{\ell}{2},n} & \equiv & \harm^\mathrm{special+}_{\ell n}  ~ ,\qquad   \lambda^-_{\ell,\frac{\ell}{2},n} \ \equiv \ \lambda^\mathrm{special+}_{\ell n}  ~ , \nn\\
\harm^-_{\ell,-\frac{\ell}{2}-1,n} & \equiv & \harm^\mathrm{special-}_{\ell n}   ~, \qquad \lambda^-_{\ell,-\frac{\ell}{2}-1,n} \ \equiv \ \lambda^\mathrm{special-}_{\ell n}   ~ .
\eea
Of course, by imposing the anti-commutation relations
\be
\{ c_{\ell m n } , c_{\ell m n}^\dagger \} \ = \ \{ d_{\ell m n} , d^\dagger_{\ell m n} \} \ = \ \delta_{\ell , \ell'} \delta_{m,m'} \delta_{n,n'}~ ,
\ee
one finds the field $\psi_\alpha$ and the conjugate momentum $\pi^\alpha = \ii \ti \psi_{\dot \alpha} \ti \sigma^{0\, \dot \alpha \alpha}$ satisfy the canonical relations
\bea
\{ \psi_\alpha (t,\vec x) , \pi^{\beta} (t,\vec x\,') \} & = & \frac{\ii}{\sqrt{-g}} \delta^{(3)}(\vec x - \vec x \,') \, {\delta_{\alpha}}^{\beta} ~,\nn\\
\{ \psi_\alpha (t,\vec x) , \psi_\beta (t,\vec x\,') \} & = & \{ \pi^\alpha (t,\vec x) , \pi^\beta (t,\vec x\, ') \} \ = \ 0~ .\label{CanAnti}
\eea
The mode expansion \eqref{fermion modes} can now be inserted into the conserved charges of section \ref{sec:Conserved Charges}, recalling that 
these have to be Weyl ordered. For example, the Hamiltonian density in \eqref{FermHamDen} becomes
\be
\mathcal H_\mathrm{fer} \ = \  \frac12 \big ((\mathcal O_f \psi) \ti \psi   -\ti \psi \mathcal O_f \psi \big) ~.   \label{WeylOrdered}
\ee
Inserting the mode expansion and integrating over the $S^3$ yields the quantized Hamiltonian
\bea
H_\mathrm{fer} & = & \frac12  \sum_{\ell=0}^\infty \sum_{n=-\frac{\ell}{2}}^{\frac{\ell}{2}} \sum_{m=-\frac{\ell}{2}-1}^{\frac{\ell}{2}} \lambda^-_{\ell m} \left(  c_{\ell m n } c^\dagger_{\ell m n}-c_{\ell m n }^\dagger c_{\ell m n} \right) \nn\\
&& - \frac12 \sum_{\ell=1}^\infty \sum_{n=-\frac{\ell}{2}}^{\frac{\ell}{2}} \sum_{m=-\frac{\ell}{2}}^{\frac{\ell}{2}-1} \lambda^+_{\ell m}\left(  d_{\ell m n} d^\dagger_{\ell m n} - d_{\ell m n }^\dagger d_{\ell m n} \right)  \label{fermHamiltonian} ~.
\eea
 
In the next section we will turn to the computation of the expectation values of these Hamiltonians, and we will show that 
the infinite sums can be evaluated with (Hurwitz) zeta function regularization in two special cases. One case is obtained for $\gag=0$, for which we can use the pairing of bosonic and fermionic eigenvalues discussed in section \ref{sec:path integral} to 
evaluate the vev of $H = H_\mathrm{bos} + H_\mathrm{fer}$.   Another case is obtained for $\beta_f = \beta_b =0$, where we will be able to evaluate the vevs of $H_\mathrm{bos}$ and
$H_\mathrm{fer}$ separately.

Thus, for simplicity in the remainder of this section we restrict to $\beta_f = \beta_b =0$. Using the mode expansions of the fields, 
and after Weyl ordering, we obtain  expressions for  the remaining conserved charges of section \ref{sec:Conserved Charges}.
For the $R$-charge, equation \eqref{JR}, this leads to
\bea
R &=&  \frac{r}{2} \sum_{\ell=0} \sum_{m,n=-\frac{\ell}{2}}^{\frac{\ell}{2}}  \big( a_{\ell m n} a^\dagger_{\ell m n} +a^\dagger_{\ell m n} a_{\ell m n}\big)  - \frac{r}{2} \sum_{\ell=0} \sum_{m,n=-\frac{\ell}{2}}^{\frac{\ell}{2}}  \big( b^\dagger_{\ell m n} b_{\ell m n} + b_{\ell m n} b^\dagger_{\ell m n} \big)  \nn\\
& & - \frac{r-1}{2} \sum_{\ell=0}^\infty \sum_{n=-\frac{\ell}{2}}^{\frac{\ell}{2}} \sum_{m=-\frac{\ell}{2}-1}^{\frac{\ell}{2}} \left(  c_{\ell m n}   c^\dagger_{\ell m n} -c^\dagger_{\ell m n}  c_{\ell m n}  \right) \nn\\
  &&  + \frac{r-1}{2}  \sum_{\ell=1}^\infty \sum_{n=-\frac{\ell}{2}}^{\frac{\ell}{2}} \sum_{m=-\frac{\ell}{2}}^{\frac{\ell}{2}-1} \left( d_{\ell m n}  d_{\ell m n}^{\dagger}- d_{\ell m n}^{\dagger} d_{\ell m n}  \right) ~ . \label{Rcharge}
\eea
For  the $J_3$ angular momentum, equation \eqref{JtJ3}, we get
\bea
J_3&=&   \frac12 \sum_{\ell=0}^\infty \sum_{m=-\frac{\ell}{2}}^{\frac{\ell}{2}} \sum_{n=-\frac{\ell}{2}}^{\frac{\ell}{2}}  m  \big( a_{\ell  mn}    a^\dagger_{\ell m n} +  a^\dagger_{\ell mn}     a_{\ell m n} \big) \nn\\
&&+  \frac12 \sum_{\ell=0}^\infty \sum_{m=-\frac{\ell}{2}}^{\frac{\ell}{2}} \sum_{n=-\frac{\ell}{2}}^{\frac{\ell}{2}}  m \big( b^\dagger_{\ell mn}   b_{\ell mn} +  b_{\ell mn}  b^\dagger_{\ell mn} \big)  \nn\\
&& - \frac12    \sum_{\ell=0}^\infty \sum_{n=-\frac{\ell}{2}}^{\frac{\ell}{2}} \sum_{m=-\frac{\ell}{2}-1}^{\frac{\ell}{2}} \left( m + \frac12 \right)  \left(  c_{\ell m n}   c_{\ell m n}^\dagger - c_{\ell m n}^\dagger     c_{\ell m n} \right) \nn\\
&&+  \frac12   \sum_{\ell=1}^\infty \sum_{n=-\frac{\ell}{2}}^{\frac{\ell}{2}} \sum_{m=-\frac{\ell}{2}}^{\frac{\ell}{2}-1} \left( m +\frac12 \right) \left(  d_{\ell m n} d^\dagger_{\ell m n } - d^\dagger_{\ell m n} d_{\ell m n } \right) ~ , \label{J3}
\eea
and finally the supercharge \eqref{Q} reads
\bea
Q &=& - \ii    \sum_{\ell=0}^\infty \sum_{n=-\frac{\ell}{2}}^{\frac{\ell}{2}} \sum_{m=-\frac{\ell}{2}}^{\frac{\ell}{2}} \sqrt{\frac{\ell}2+ m+1} \   a^\dagger_{\ell mn}  c_{\ell m n}  \nn\\
&& -\ii  \sum_{\ell=1}^\infty  \sum_{n=-\frac{\ell}{2}}^{\frac{\ell}{2}}\sum_{m=-\frac{\ell}{2}}^{\frac{\ell}{2}-1}    (-1)^{-m-n}   \sqrt{\frac{\ell}{2}- m} \  b_{\ell,- m,- n} d_{\ell m n}^{\dagger} ~ .  \label{supercharge}
\eea

By direct computation, one can now verify the following commutation relations
\bea
\left[H , Q  \right] &=& -\frac{\gag}{\sfrad} Q  ~ , \qquad \left[ R, Q  \right] \ = \   Q  ~, \qquad \left[J_3 , Q  \right] \ = \  -\frac12 Q ~ ,
\label{commuts}
\eea
where we restored the radius $\sfrad$ of the $S^3$. Note that the Hamiltonian commutes with $Q$ only for $\gag =0$, which from  equation \eqref{thepair} is the value required for the pairing of eigenvalues. By conjugating equation \eqref{supercharge}, one can further verify that
\be
\frac{\sfrad}{2} \{Q,Q^\dagger \} \ = \ :H + \frac{1}{\sfrad} (1+\gag) R + \frac{2}{\sfrad} J_3:  ~ ,   \label{QQdagger}
\ee
where $:~ :$ denotes normal ordering. 

Let us  set  $\sfrad=1$ in (\ref{QQdagger}) and  (\ref{commuts})  and comment on the special values of the parameter $\gag$ discussed in the literature. 
Setting  $\gag=0$ we have 
\bea
\frac{1}{2} \{Q,Q^\dagger \} & = & : H +  R + 2 J_3 :   ~ , \nn\\
{[}H,Q] & = & 0~,   
\eea
corresponding\footnote{Here and below, the equations correspond 
up to convention dependent signs of $R$ and $J_3$, as well a possible factor $\sqrt 2$ in the supercharge $Q$, descending from the definition of the 
 supersymmetry variations.} to the relations in  eq. (5.9) of  \cite{Festuccia:2011ws}, where $H|_{\gag=0}$ coincides with  $H$  in that reference. 
 For this reason we refer to $H|_{\gag=0}\equiv H_\mathrm{susy}$ as the BPS  Hamiltonian. 

Setting $\gag=1/2$  we have
\bea
\frac{1}{2} \{Q,Q^\dagger \} & = & : H +  \frac{3}{2}R + 2 J_3 :   ~ , \nn\\
{[}H,Q] & = & -\frac{1}{2} Q~,   
\eea
which coincide for example with eq.  in (7) of \cite{Romelsberger:2007ec} as
well as with eq. (6.11) in \cite{Festuccia:2011ws}, where $H|_{\gag=1/2}$ corresponds to $\Delta$ in the latter reference. 

Finally, setting  $\gag=-1$ we have 
\bea
\frac{1}{2} \{Q,Q^\dagger \} & = & : H +  2 J_3:  ~ ,   \nn\\
{[}H,Q] & = &  Q~,   
\eea
corresponding to eq. (5.6) of \cite{Festuccia:2011ws}, where  $H|_{\gag=-1}$ corresponds to $P_0$ in that reference. 

Although these commutation relations are here written for the chiral multiplet, it is straightforward to verify that they hold also for the vector multiplet, and hence for the 
total  $H_\mathrm{tot}= H+ H_\mathrm{vec}$, and similarly for the other operators.  It was  noticed in  \cite{Cassani:2014zwa} that these may be formally derived from the abstract supersymmetry 
algebra of new minimal supergravity.

\subsection{Casimir energy}

We are now ready to compute the vacuum expectation value of the Hamiltonian. This yields infinite sums which we  regularize using the  zeta function method. Thus, for an
 operator $A$, we define its vacuum expectation value as 
 \bea
   \langle  A \rangle & \equiv  & \lim_{s \to -1} \zeta_A (s) ~,
 \eea
where, denoting with  $\lambda_n^A$ the set of all the eigenvalues (here $n$ is a multi-index) 
of $A$ and with $d^A_n$ their degeneracies, the generalised zeta function is defined as
\bea
 \zeta_A (s)  & = &  \mathrm{Tr} \, A^{-s} \ = \    \sum_{n} d^A_n (\lambda^A_n)^{-s} ~.
 \label{multsit}
\eea
Notice that if $A=B+C$, with corresponding eigenvalues denoted as  $\lambda^B_n$ and $\lambda^C_n$, then
\be
\lim_{s \to -1} \left(\sum_{n}  (\lambda^B_n)^{-s} + \sum_{n }  (\lambda_n^C)^{-s} \right) \  \neq \ \lim_{s \to -1} \sum_{n}  (\lambda_n^B + \lambda^C_n)^{-s} ~ .
\label{different}
\ee
This lack of additivity is related to the lack of associativity of functional determinants, 
 $\det (BC)\neq \det (B) \cdot \det (C)$, which is known as  ``multiplicative anomaly''. See e.g.   \cite{Elizalde:2012zza}.

In the present context, we use the following prescription for dealing with the infinite sums: for each given operator, we sum independently the  eigenvalues corresponding to
every different field.  In particular, we define the vev of each operator as the sum of the vevs of the terms containing 
the fields $\phi$, $\psi$, ${\cal A}$, and $\lambda$, respectively.  Therefore, for example, 
\bea
\langle H \rangle & \equiv  & \langle H_\mathrm{bos} \rangle + \langle H_\mathrm{fer} \rangle~, 
\label{defhvev}
\eea
and similarly for $R$, $J_3$, and $Q$. This recipe is  in accordance with \cite{Dowker:1998tb} and
yields to  the supersymmetric Casimir energy computed in \cite{Assel:2014paa}.

The vevs of the scalar and fermion Hamiltonians, \eqref{scalarHam} and \eqref{fermHamiltonian}, are
\bea
\langle H_\mathrm{bos} \rangle & = &  \lim_{s\to -1} \left[\frac{1}{2} \sum_{\ell=0}^\infty \sum_{m,n=-\frac{\ell}{2}}^{\frac{\ell}{2}} (\omega^{+}_{\ell m })^{-s} + \frac{1}{2} \sum_{\ell=0}^\infty \sum_{m,n=-\frac{\ell}{2}}^{\frac{\ell}{2}} (\omega^{-}_{\ell m})^{-s}  \right]~,  \label{BosonicCasimir} \nn\\[2mm]
\langle H_\mathrm{fer} \rangle & = &  \lim_{s\to -1} \left[ \frac12  \sum_{\ell=0}^\infty \sum_{n=-\frac{\ell}{2}}^{\frac{\ell}{2}} \sum_{m=-\frac{\ell}{2}-1}^{\frac{\ell}{2}} (\lambda^-_{\ell m })^{-s} - \frac12 \sum_{\ell=1}^\infty \sum_{n=-\frac{\ell}{2}}^{\frac{\ell}{2}} \sum_{m=-\frac{\ell}{2}}^{\frac{\ell}{2}-1} (\lambda^+_{\ell m} )^{-s} \right]~ ,\label{FermionCasimir}
\eea
 respectively. However, due to the square roots appearing in both sets of eigenvalues $\omega^\pm_{\ell m}$ and $\lambda^\pm_{\ell m}$, the vevs in \eqref{FermionCasimir} cannot in general be separately regularized with any\footnote{E.g. Hurwitz, Barnes, Shintani, Epstein zeta functions.} zeta function and written in closed form.

In the special case $\gag=0$, we can take advantage of the pairing as discussed in section \ref{sec:path integral} to compute the vev of the Hamiltonian of the 
chiral multiplet, $H=H_\mathrm{bos} + H_\mathrm{fer}$. Thus, setting $\gag=0$ one has
\be
\omega^+_{\ell m}  \ = \ - \lambda^-_{\ell m}  ~, \qquad  \omega^-_{\ell, -m} \ =  \ \lambda^+_{\ell m}  ~ , \qquad \mathrm{for}\quad  \ell \ge 1~, \quad -\frac{\ell}{2} \le m \le \frac{\ell}{2}-1 ~ . \label{paired}
\ee
The eigenvalues not included in equation \eqref{paired} are the ``special'' fermion eigenvalues, which we can write as
\be
\lambda^\mathrm{special\pm}_\ell \ = \ -\frac{1}{2} (\ell+1) \pm \beta_f - \deltab ~, \qquad \ell \ge 0 ~ ,
\ee
and the ``unpaired'' bosonic eigenvalues
\be
\omega^+_{\ell,\frac{\ell}{2}} \ = \ \frac{1}{2} (\ell+1)  - \beta_f + \deltab ~ , \qquad \omega^-_{\ell,-\frac{\ell}{2}} \ = \ \frac{1}{2} (\ell+1)  - \beta_f - \deltab ~   , \qquad \ell \ge0 ~ .
\ee
Here we used  that $\alpha_f<0$ and assumed $\beta_f \le - \frac{\alpha_f}{2}$ in order simplify the square roots in $\omega^+_{\ell, \frac{\ell}{2}}$ and $\omega^-_{\ell , - \frac{\ell}{2}}$. 
Due to the pairing, equation \eqref{paired}, all eigenvalues containing square roots exactly cancel against each other  in (\ref{defhvev}), and we are left with
\bea
\langle H\rangle_{\gag =0 }  & =& \lim_{s\to -1} \Bigg[\frac{1}{2} \sum_{\ell=0}^\infty \sum_{n=-\frac{\ell}{2}}^{\frac{\ell}{2}} (\omega^{+}_{\ell ,\frac{\ell}{2} })^{-s} + \frac{1}{2} \sum_{\ell=0}^\infty \sum_{n=-\frac{\ell}{2}}^{\frac{\ell}{2}} (\omega^{-}_{\ell , - \frac{\ell}{2} })^{-s} \nn\\
&&  \qquad ~ + \frac12  \sum_{\ell=0}^\infty \sum_{n=-\frac{\ell}{2}}^{\frac{\ell}{2}} (\lambda^\mathrm{special +}_{\ell})^{-s} + \frac12 \sum_{\ell=0}^\infty \sum_{n=-\frac{\ell}{2}}^{\frac{\ell}{2}}  (\lambda^\mathrm{special-}_{\ell } )^{-s}\Bigg]\nn\\
& =&  \lim_{s\to -1}\left[ \frac{1}{4} \sum_{k=1}^\infty k \left(  k - 2(\beta_f +\deltab ) \right)^{-s}   - \frac{1}{4}\sum_{k=1}^\infty k \left(k  + 2(\beta_f+ \deltab)  \right)^{-s} \right]\nn\\[2mm]
&=& \frac{1}{12}(\beta_f + \deltab) \left(1- 8 (\beta_f + \deltab)^2\right)~ .  \label{pairedresult}
\eea
 Notice that the first and third term in the first line further exactly cancelled and in the last step we regularized \emph{separately} the two remaining sums using the Hurwitz zeta function.
See appendix \ref{hurwapp}.
To summarize, since for $\gag=0$ one has $2(\beta_f + \deltab)= 1-r$, the vev of  the Hamiltonian of a chiral multiplet with $R$-charge $r$ is
\be
\langle H \rangle_{\gag =0 }   \ = \ \frac{1}{12 \sfrad}(1-r) \left(1- 2 (1-r)^2\right)~,
\ee
where we restored $\sfrad$. This result is valid for any value of $r$, $\kappa$, $\epsilon$.  Notice that if we were to combine the two sums in the middle line of (\ref{pairedresult}),  before regularization, we would get a different result. 

It is now simple to combine this with the contributions of the fields in the vector multiplet, and recover 
the supersymmetric Casimir energy $E_\mathrm{susy}$ of section \ref{sec:path integral}. 
The Casimir energy of the gauge field does not depend on any of our parameters and is simply given by the result for an Abelian gauge field 
$\langle H_\mathrm{gauge}\rangle  = \frac{11}{120 \sfrad}$ (see e.g. \cite{Cappelli:1988vw}) multiplied by the dimension of the gauge group $N_v$. 
 For the gaugino $\lambda$, the Casimir energy is computed as for the fermion $\psi$ above, 
but using the operator $\mathcal O_f^\mathrm{vec}$ in equation \eqref{Ofvector}, giving simply  $\langle H_\mathrm{gaugino} \rangle   = - \frac{1}{120\sfrad}$, again multiplied by 
$N_v$.  Adding everything together, we obtain
\bea
\langle H_\mathrm{tot} \rangle_{\gag =0} & = &    \frac{1}{12 \sfrad} \left( 2 \mathrm{tr}\Rch^3 -  \mathrm{tr}\Rch   \right) \ = \ \frac{4}{27 \sfrad} (\mathbf a+3 \mathbf c)  \ = \ \frac{1}{\sfrad} E_\mathrm{susy}~.
\label{essu}
\eea
This result is valid for arbitrary values of $\kappa$ and $\epsilon$, in agreement\footnote{The quantity $E_\mathrm{susy}$ defined in \cite{Assel:2014paa} is dimensionless. Therefore, if writing 
the radius of the three-sphere explicitly, this has 
to be compared with the dimensionless combination $\sfrad \langle H_\mathrm{tot} \rangle_{\gag=0}$.} with (\ref{Erecover}).
Indeed this is exactly the same BPS Hamiltonian defining the path integral, and therefore the free field result should have agreed with the localization result, that is valid for any value of the couplings.

Next, we consider the special case $\beta_b=\beta_f=0$. This corresponds to setting $\kappa=-1$ and $\epsilon=1$, but leaving arbitrary $\gag$. 
In this case, both sums in  \eqref{BosonicCasimir} can be separately regularized using Hurwitz zeta function, as the square roots in $\omega^\pm_{\ell m}$ and $\lambda^\pm_{\ell m}$ are absent, namely
\be
\omega^+_{\ell }  \ = \ \frac{1}{2} (\ell+2 - r (1+\gag) ) ~, \qquad \omega^-_{\ell } \ = \ \frac{1}{2} (\ell + r(1+\gag) ) ~ , \label{omega} 
\ee
and 
\be
\lambda^-_{\ell} \ = \ \lambda^\mathrm{special \pm}_{\ell } \ = \ -\frac{1}{2} (\ell +2-r( 1 +\gag) +\gag )  ~, \qquad \lambda^+_{\ell } \ = \ \frac{1}{2} (\ell+r(1+\gag ) -\gag)  ~ , \label{beta=0eigen}
\ee
where we dropped the subscript $m$, as this quantum number becomes degenerate. 
Thus, regularizing the sums  as described at the beginning of this subsection using the Hurwitz zeta function, we obtain the finite Casimir energies
\bea
\langle H_\mathrm{bos} \rangle & = &  \frac{1}{240} \left[1-10 \big( r (1+\gag) -1 \big)^4\right]  \label{BosonicCasimir2} ~ , 
\eea
and 
\bea
\langle H_\mathrm{fer} \rangle & = &   \frac{1}{240} \Big[ 10 (\gag+1)^3 \left(1+\gag \right) (r-1)^4\nn\\[2mm]
&&+20 (\gag+1)^3 (r-1)^3- 10 (\gag+1) (r-1)-1\Big] ~ \label{beta0} .
\eea

Adding \eqref{BosonicCasimir2} and \eqref{beta0}, we obtain the Casimir energy of a chiral multiplet with $R$-charge $r$
\bea
\langle H  \rangle & = & - \frac{1}{24} \Big[ \gag^4 +2 (\gag+1)^3 (2 \gag-1) (r-1)^3 \nn\\
&& + 6 \gag^2(\gag+1)^2  (r-1)^2 +(\gag+1)\left(4 \gag^3+1\right)   (r-1)  \Big] ~ .  \label{vevHbeta=0}
\eea
This generalizes straightforwardly to an arbitrary number of chiral multiplets.  As before,  we can include easily an arbitrary number $N_v$ of vector multiplets as well.  In this case, for the gaugino, the Casimir energy  can be obtained by formally setting 
$r = \frac{2 \gag}{1+\gag}$ \ in equation \eqref{beta0}, and reads
\be
\langle H_\mathrm{gaugino} \rangle  \ = \ \frac{1}{240} \left(10 \left(\gag^4-2 \gag^3+\gag\right)-1\right) . 
\ee

Combining these results, we find that (for $\kappa=-1$, $\epsilon=1$) using our regularization, the Casimir energy of a supersymmetric gauge theory with 
$N_v$ vector multiplets and $N_\chi$ chiral multiplets with $R$-charges $r_I$ is given by the  following expression
\bea
 \langle H_\mathrm{tot} \rangle & = & \frac{N_v}{12\sfrad} \left(\gag^4-2 \gag^3+\gag+1\right)   -\frac{1}{12\sfrad} \sum_{I=1}^{N_\chi} \Big (\gag^4+\left(4 \gag^3+1\right) (\gag+1)  (r_I-1)\nn\\
&&+6  \gag^2 (\gag+1)^2 (r_I-1)^2  +2 (2 \gag-1)  (\gag+1)^3(r_I-1)^3 \Big)  ~ ,  \label{FullCasimir}
\eea
where we restored the radius $\sfrad$ of the three-sphere. 
Setting   $\gag=0$ as in \cite{Assel:2014paa}, 
we see that \eqref{FullCasimir} reduce to
\be
\langle H_\mathrm{tot} \rangle_{\gag=0} \ = \ \frac{4}{27 \sfrad} (\mathbf a+3 \mathbf c) ~,
\ee
in agreement with (\ref{essu}).

In general, however, equation \eqref{FullCasimir} cannot be written as a linear combination of $\mathbf a$ and $\mathbf c$. 
In the special case  $\gag=1/2$ and $r_I=2/3$, corresponding to the usual  conformally coupled scalars, Weyl spinors, and gauge fields, the Casimir energy \eqref{FullCasimir} reduces to
\be
\langle H_\mathrm{tot} \rangle_{\gag=\frac12, r_I=\frac23} \ = \  \frac{1}{192 \sfrad} (21 N_v+5 N_\chi) \ = \ \frac{1}{4 \sfrad} (\mathbf a+2 \mathbf c) ~ , \label{FreeCasimir}
\ee
in accordance with standard zeta function computations (see e.g. \cite{Marino:2011nm}). 
In particular,  notice that for theories with $N_\chi = 3 N_v$ (so that $\mathbf a=\mathbf c$), such as  ${\cal N}=4$ super-Yang Mills,
this becomes simply $\frac{3}{4 \sfrad} \mathbf a$. However, the agreement with the CFT result of  \cite{Herzog:2013ed} for the Casimir energy is accidental \cite{Cappelli:1988vw,Herzog:2013ed}. Finally, we note that for $\gag =-1$,  the Casimir energy is independent of the $R$-charges and reads
\bea
 \langle H_\mathrm{tot} \rangle_{\gag=-1} & = & \frac{1}{12 \sfrad}  \left( 3 N_v  -N_\chi \right)   ~ .
  \label{peculiar}
\eea
This is simply because in this case $A=0$, and therefore the Lagrangian does not depend on the $R$-charges, as we observed at the end of section
\ref{sec:background}.

We can also  compute the vev\footnote{The  vev of the supercharge $Q$ is zero. For the chiral multiplet this follows 
from (\ref{Q}) or its mode expansion (\ref{supercharge}). Similarly, for the vector multiplet it follows from the explicit expression of $Q_\mathrm{vec}$, which is $
Q_\mathrm{vec} \ = \  \frac{\ii}{2}  \int \sqrt{g_3} \dd^3 x  \  \zeta \sigma^{\mu} \tilde \sigma^\nu \sigma^0 \ti \lambda \mathcal F_{\mu \nu}  $.} 
of the $R$-charge operator $R$, by using  the regularization  described above.
For a single chiral multiplet of $R$-charge $r$ we get 
$\langle R \rangle   =    \frac1{12} (r-1)$,  
where only the fermion $\psi$ contributes, while for an (Abelian)  vector multiplet 
we get  $\langle R_\mathrm{vec} \rangle \ = \ \frac{1}{12}$,
where again only the gaugino contributes. Thus, for the total $R$-charge operator $R_\mathrm{tot}= R+R_\mathrm{vec}$, we find 
\bea
\langle R_\mathrm{tot} \rangle  & =& \frac{4}{3}(\mathbf a-\mathbf c) ~ .
\eea

The results discussed in this section   rely on the fact  that the operators we are using are not normal ordered. See also  \cite{McKenzieSmith:2000vz} for a similar discussion.

\section{Conclusions}

\label{concl:section}

The main purpose of this note was to  examine aspects of the supersymmetric Casimir energy of ${\cal N}=1$ gauge theories introduced in \cite{Assel:2014paa}, 
in the simplest case where the partition function depends only on one fugacity.  It has been argued recently 
in  \cite{Assel:2014tba} that this quantity does not suffer from ambiguities, and therefore it should be  a good physical observable.

Firstly, by revisiting the localization computation in \cite{Assel:2014paa},
 we have verified explicitly  that its value does not depend on the choice of the parameter $\kappa$, characterizing the background fields $A$ and $V$, as expected. 
Secondly,  we reproduced it by evaluating the expectation value of the BPS Hamiltonian that appears in the definition of the supersymmetric index, as anticipated in 
\cite{Kim:2012ava}. Our computations also clarify the relation of the supersymmetric Casimir energy with the Casimir energy of free conformal fields theories, 
demonstrating that these two quantities arise as the expectation values of two \emph{different} Hamiltonians, evaluated using the \emph{same} zeta function 
regularization method.

An obvious extension of this note is to compute the vev of 
the BPS Hamiltonian in the more general case of supersymmetric theories on a Hopf surface with two complex structure parameters, and to check that this agrees with the general form of the supersymmetric Casimir energy  \cite{Assel:2014paa}. 
It should  also be rewarding to study this problem from the viewpoint of  
\cite{Brown:1977sj,Page:1982fm,Cappelli:1988vw,Herzog:2013ed}, 
 or using the effective action approach of \cite{DiPietro:2014bca}. 

 The discrepancy between the   supersymmetric Casimir  energy and the renormalized on-shell action in AdS$_5$,  pointed out in  \cite{Cassani:2014zwa},  remains unexplained.  
 We think  it  is imperative to elucidate  how to reproduce   the supersymmetric Casimir energy from a holographic computation.

\subsection*{Acknowledgements}
We are supported by the ERC Starting Grant N. 304806, ``The Gauge/Gravity Duality and Geometry in String Theory''.
We thank B. Assel,  D. Cassani, L. Di Pietro, and Z.  Komargodski for discussions on related topics. We are grateful to 
B. Assel and D. Cassani for  comments on  a draft.

\appendix

\section{Spherical harmonics}
\label{App: Harmonics}

\subsection{Scalar spherical harmonics}

In this appendix we give some details on the scalar and spinor spherical harmonics on the three-sphere, following 
\cite{Sen:1985dc,Cutkosky:1983jd}. We can obtain the metric on the unit three-sphere by considering a parametrization on $\mathbb R^4 \simeq \mathbb C^2$ with the metric,
\be
\dd s_{\mathbb C^2 }^2 \  = \ \dd u \dd \bar u + \dd v \dd \bar v ~.
\ee
The three-sphere of unit radius is then defined by
\be
u \bar u + v \bar v \ = \ 1~.
\ee
The isometry group is $SO(4) \simeq SU(2)_L \times SU(2)_R$, with generators\footnote{In the main text, we use the operators $L^L_a$, but drop the superscript $L$.} $L^L_a$ and $L^R_a$, with $a=1,2,3$, satisfying
\be
\left[ L^L_a , L^L_b \right] \ =  \ \ii \epsilon_{abc} L^L_c ~, \qquad \left[ L^R_a , L^R_b \right] \ =  \ \ii \epsilon_{abc} L^R_c ~, \qquad  \left[ L^L_a, L^R_b \right] \ = \ 0~. 
\ee
As usual, we define raising and lowering operators, 
\be
L_\pm^L \ = \ L_1^L \pm \ii L_2^L ~, \qquad L_\pm^R \ = \ L_1^R \pm \ii L_2^R~.
\ee
In the $(u,v)$-coordinates, these are represented by
\bea
L_+^L  & = & - u \de_{\bar v} + v \de_{\bar u} ~, \quad L^L_- \ = \ \bar u \de_v - \bar v \de_u ~,  \nn\\
L_+^R  & = & -u \de_{ v} + \bar v \de_{\bar u} ~, \quad L^R_- \ = \ \bar u \de_{\bar v} - v \de_u ~, \label{leftright} 
\eea
while 
\bea
L^L_3 & = & \frac{1}{2} \left(u \de_u + v \de_v - \bar u \de_{\bar u} - \bar v \de_{\bar v} \right)  ~, \qquad L^R_3 \ = \ \frac{1}{2} \left(u \de_u - v \de_v - \bar u \de_{\bar u} + \bar v \de_{\bar v} \right) ~. \  \ 
\eea
In terms of these operators, the scalar Laplacian is
\be
- \nabla_i \nabla^i \ =  \ 4 L^L_a L^L_a \ = \ 4 L^R_a L^R_a ~.
\ee
The spherical harmonics $Y_\ell^{mn}$ are constructed starting from the highest weight state,
\be
Y_\ell^{\frac{\ell}{2}\frac{\ell}{2}} \ = \ \sqrt{\frac{\ell+1}{2\pi^2}} \ u^\ell~,
\ee
which is annihilated by the raising operators $L^L_+$ and $L^R_+$.  The number $m$ ($n$) can be lowered by $L_-^L$ ($L^R_-$), so that 
\be
Y_{\ell}^{m n } \ \propto \ \left( L^L_-\right)^{\frac{\ell}{2} - m}  \left( L^R_-\right)^{\frac{\ell}{2} - n} Y_\ell^{\frac{\ell}{2}\frac{\ell}{2}}~,
\ee
and take values $-\frac{\ell}{2}  \le m,n \le \frac{\ell}{2}$. The spherical harmonics are eigenfunctions of the operator $\mathcal O_b$, equation \eqref{bloppy}, 
\bea
\mathcal O_b Y^{mn}_\ell \ =  \ E^2_b Y^{mn}_\ell ~, \qquad E^2_b \ = \ \frac{\alphab}{2}\ell  (\ell +2) + 2 \betab \mkap + \gammab ~,
\eea
and also
\be
L^L_3 Y^{mn}_\ell \ = \ m Y^{mn}_\ell ~ ,\qquad L^R_3 Y^{mn}_\ell \ = \ n Y^{mn}_\ell~. 
\ee
We normalize the spherical harmonics as \cite{Cutkosky:1983jd} 
\be
Y_\ell^{\frac{\ell}{2}-a , \frac{\ell}{2}-b} \ = \ N_{\ell a b} \sum_{k} \frac{(-u)^{\ell+k-a-b}\bar u^k v^{b-k} \bar v^{a-k} }{k! (\ell +k-a-b)! (a-k)!(b-k)!}~, \label{lower}
\ee
where the sum is over all integer values of $k$ for which the exponents are non-negative, and
\be
N_{\ell a b} \ = \ \sqrt{\frac{(\ell+1)a!  b!(\ell-a)! (\ell-b)!}{2\pi^2}} ~ .
\ee

Now specifically taking
$u= \ii \sin\frac{\theta}{2} \ex^{\ii (\varphi-\varsigma  )/2}$ and $v = \cos\frac{\theta}{2} \ex^{-\ii (\varphi+\varsigma)/2}$, one finds the metric
\be
\dd s_{S^3}^2 \ =  \ \frac{1}{4} \left( \dd \theta^2 + \sin^2 \theta \dd \varphi^2 +(\dd \varsigma+ \cos \theta \dd \varphi)^2 \right) ~,
\ee
and
\be
L_3^L \ = \ \ii \de_\varsigma ~\ \qquad L_3^R \ = \ -\ii \de_\varphi~.
\ee
With the above normalization, the spherical harmonics satisfy
\be
\int \sqrt{g_3} \dd^3 x Y^{mn}_{\ell} \left( Y^{m'n'}_{\ell'} \right)^\ast \ = \ \delta_{\ell,\ell'}\delta^{m,m'}\delta^{n,n'}~, \label{orthonormal}
\ee
 and 
\be
\left(Y^{mn}_\ell \right)^\ast \ = \ (-1)^{m+n} Y_\ell^{-m,-n} ~, \label{complex}
\ee
as well as the completeness relation,
\bea
\sum_{\ell,m,n} Y_{\ell}^{mn} (\theta,\varphi , \varsigma) \left( Y^{mn}_\ell (\theta ',\varphi', \varsigma') \right)^\ast  &  =  &\frac{1}{\sin \theta} \delta^{(3)}( \vec x - \vec x \, ')~, \label{completeness}
\eea
where $\delta^{(3)}( \vec x - \vec x \, ') = \delta(\theta - \theta') \delta(\varphi-\varphi') \delta( \varsigma  -\varsigma ')$.

\subsection{Spinor spherical harmonics}
\label{sec:spinor harm}

The spinor spherical harmonics can be constructed from the scalar harmonics. These are eigenspinors of the operator
\be
\fop \ =  \ 2 \alpha_f \vec{L}\cdot \vec{S} + 2\beta_f S_3 + \gamma_f ~ ,
\ee 
where $L_a$ are the left-invariant operators of the previous subsection, and $S_a = \frac{\gamma_a}{2}$, where $\gamma_a$ are the Pauli matrices.
For $\beta_f=0$, the spinor spherical harmonics can be constructed as \cite{Sen:1985dc}
\be
S^\pm_{\ell m n} \ = \  \left( \begin{array}{c} \cos \nu ^\pm_{\ell m } \, Y^{mn}_\ell \\ \sin\nu^\pm_{\ell m}  \, Y^{m+1,n}_\ell \end{array} \right)~,  \label{sphericalharmonicsbeta=0}
\ee
where
\be
\sin \nu^\pm_{\ell m} \ = \ \mp \sqrt{\frac{\ell +1 \pm (2m+1)}{2  (\ell+1)} } ~ , \qquad \cos \nu^\pm_{\ell m} \ = \  \sqrt{\frac{\ell +1 \mp (2m+1)}{2  (\ell+1)} }  ~ .
\ee
For $S^+_{\ell m n }$, one has $\ell \ge1$ and $ -\frac{\ell}{2} \le m \le \frac{\ell}{2}-1$, while for $S^-_{\ell m n} $ one has $\ell \ge 0$ and $ -\frac{\ell}{2}-1 \le m \le \frac{\ell}{2}$. In both cases $-\frac{\ell}{2} \le n \le \frac{\ell}{2}$. The spinor spherical harmonics satisfy the completeness relation 
\be
\sum_{m,n} S^\pm_{\ell mn \, \alpha} (x) \left( S^\pm_{\ell m n }(x) \right)^{\dagger}_{ \dot\alpha} \ = \ \frac1{4\pi^2} n^\pm_\ell \bfone_{\alpha \dot \alpha}  ~, 
\ee
with $n^+_\ell =\ell(\ell+1) $ and $n^-_\ell = (\ell+2)(\ell+1)$. Further, using the properties of $Y^{mn}_{\ell}$, one can show the identities
\be
\sum_{\ell, m, n} \left[ S^+_{\ell m n \, \alpha} (x) \left( S^+_{\ell mn } (x')\right)^{\dagger}_{ \dot \alpha}  + S^-_{\ell m n \, \alpha} (x) \left( S^-_{\ell  mn} (x')\right)^{\dagger }_{\dot \alpha}  \right] \ = \ \frac{1}{\sin\theta} \delta^{(3)}( \vec x - \vec x \, ')\, {\bfone}_{\alpha \dot \alpha} \label{sumidentity}~,
\ee
and
\bea
\int \dd^3x \sqrt{g_3} \ S^\pm_{\ell m n \, \alpha} (x) \left(  S^{\pm'}_{\ell 'm'n' } (x)  \right)^{\dagger}_{\dot \alpha} \bfone^{\alpha \dot \alpha} & = & \delta_{\ell,\ell'}\delta_{m,m'}\delta_{n,n'}\delta^{\pm,\pm'}~,  \label{spinorintegral}
\eea
where the integral is on the unit three-sphere. Using that
\bea
L_+ Y_\ell^{mn}  & = & \frac12 \sqrt{\ell(\ell+2) -4m(m+1) } Y^{m+1,n}_\ell~, \nn \\
L_- Y_\ell^{m+1,n} & = & \frac12 \sqrt{\ell (\ell+2)-4m(m+1)} Y^{mn}_\ell    ~ , 
\eea
one can verify that
\be
\mathcal O_f S^\pm_{\ell m n } \ = \ \lambda^\pm_{\ell } S^\pm _{\ell m n } ~,
\ee
with 
\be
\lambda^+_{\ell} \ = \ -\frac{\alpha_f}{2} (\ell+2 ) + \gamma_f~, \qquad \lambda^-_\ell  \ =  \ \frac{\alpha_f}{2} \ell + \gamma_f ~ .
\ee
When $\beta_f \neq0$, the spinor spherical harmonics given by \eqref{sphericalharmonicsbeta=0} are not eigenspinors of the operator $\mathcal O_f$, except the special cases
\bea
\harm^\mathrm{special+}_{\ell n} & \equiv & S^-_{\ell,\frac{\ell}{2},n} \  = \ \left(\begin{array}{c}  Y^{\frac{\ell}{2},n}_\ell \\ 0  \end{array}   \right) ~, \qquad \mathbf  \harm^\mathrm{special-}_{\ell n} \ \equiv \ S^-_{\ell,-\frac{\ell}{2}-1,n} \  = \ \left(\begin{array}{c} 0 \\  Y^{-\frac{\ell}{2},n}_\ell   \end{array}   \right) ~ ,\nn\\
\mathcal O_f  \harm^\mathrm{special \pm}_{\ell n} & = & \lambda^\mathrm{special \pm}_\ell \harm^\mathrm{special\pm}_{\ell n}  ~ , \qquad \lambda^\mathrm{special \pm}_\ell \ = \ \left(\frac{\alpha_f}{2} \ell \pm \beta_f +\gamma_f \right) ~. 
\eea 
For the generic harmonics, the eigenspinors of $\mathcal O_f$ for general $\beta_f$ are obtained by an $SO(2)$ rotation,
\be
\left(\begin{array}{c} \harm^+_{\ell m n} \\ \harm^-_{\ell m n} \end{array}  \right) \ \equiv \  \left(\begin{array}{c c} \mathcal R_{11} & \mathcal R_{12} \\ \mathcal R_{21} & \mathcal R_{22} \end{array} \right) \left( \begin{array}{c} S^+_{\ell m n } \\ S^-_{\ell m n } \end{array} \right) ~ .
\ee
The rotation matrix is given by
\bea
\mathcal R_{12} & = & \mathcal R_{11}  \frac{ (\frac{\alpha_f}{2}  (\ell +2)+\lambda^+_{\ell m} - \beta_f -\gamma_f )}{  (\frac{\alpha_f}{2}  \ell- \lambda^+_{\ell m} + \beta_f +\gamma_f )}  \frac{\cos\nu^+_{\ell m}}{\cos\nu^-_{\ell m}}  ~ , \nn\\
\mathcal R_{21} & = & \mathcal R_{22}\frac{ (-\frac{\alpha_f}{2}  \ell + \lambda^-_{\ell m}- \beta_f -\gamma_f )}{  (-\frac{\alpha_f}{2}  (\ell +2) - \lambda^-_{\ell m}+ \beta_f +\gamma_f)}  \frac{\cos\nu^-_{\ell m}}{\cos\nu^+_{\ell m}} ~ ,
\eea
with 
\be
\lambda^\pm_{\ell m}  \ = \ -\frac{\alpha_f}{2} + \gamma_f \pm \sqrt{\frac{\alpha_f^2}{4} (\ell+1)^2 + \alpha_f \beta_f (1+ 2 m ) +\beta_f^2}  ~ .
\ee
Requiring the matrix to be $SO(2)$ fixes all the $\mathcal R_{ij}$, with a choice of overall sign fixed by requiring the matrix to be the identity matrix for $\beta_f=0$. We then have
\be
\mathcal O_f \harm^\pm_{\ell m n } \ =\  \lambda^\pm_{\ell m} \harm^\pm_{\ell m n }  ~ ,
\ee
for $\ell \ge 1$, $-\frac{\ell}{2} \le m \le -\frac{\ell}{2}-1 $, and $-\frac{\ell}{2} \le m \le \frac{\ell}{2}$.

\section{Hurwitz zeta function}
\label{hurwapp}

In this appendix we include the  definition of the Hurwitz zeta function and some useful properties.
This is  defined as the analytic continuation to complex $s\neq 1$, of the following series
\be
\zeta_H(s,a) \, = \, \sum_{n=0}^\infty \frac{1}{(n+a)^s}~,
\label{defzeta}
\ee
which is  convergent for any Re$(s)>1$. Notice that 
\be
\zeta_H (s,1) \, = \, \zeta (s)~,
\ee
corresponds to the Riemann zeta function. 

For $s=-k$, where $k=0,1,2,\dots$, the Hurwitz zeta function reduces to the Bernoulli polynomials
\bea
\zeta_H (-k,a) & = & - \frac{B_{k+1}(a)}{k+1}~,
\label{hurwitzbern}
\eea
defined as
\bea
B_k (a)  & = & \sum_{n=0}^k  {k \choose n}  b_{k-n} a^n\, ,
\label{defber}
\eea
where  $b_n$ are the Bernoulli numbers. The first few ones read 
\bea
B_0 (a) & = & 1~,\nn\\
B_1 (a) & = & a-\frac{1}{2}~,\nn\\
B_2 (a) & = & a^2 - a +\frac{1}{6}~,\nn \\
B_3 (a) & = & a^3 - \frac{3}{2}a^2 +\frac{1}{2}a~,\nn\\
B_4 (a) & = & a^4 - 2a^3 + a^2 -\frac{1}{30}~.
\eea
 The following formulas used in the text  are easily proved
\bea
\sum_{k=1}^\infty \frac{k}{(k +a )^{s}}  & = & \zeta_H (s-1,a) - a \,  \zeta_H (s,a)~,
\eea 
and
\bea
\sum_{k=1}^\infty \frac{k (k+1)}{(k+a)^s} & = & 
\zeta_H(s-2,a)+ (1-2a)\zeta_H(s-1,a)+a(a-1)\zeta_H (s,a)~ .
\eea

\section{Energy-momentum tensor and other currents}

\label{emten:appendix}

In this appendix  we provide explicit expressions for the energy-momentum tensor and other currents obtained from the (quadratic) chiral multiplet Lagrangian 
(\ref{lagdef}).    Denoting with $S$ the corresponding action, the energy-momentum tensor is defined as 
\be
T_{\mu \nu} \  = \   \frac{-2}{\sqrt{-g}} \frac{\delta S}{\delta g^{\mu\nu}}~.
\ee
A straightforward but tedious computation yields
\bea
T_{\mu\nu} & = & (2 \delta_{(\mu}^\rho \delta_{\nu)}^\lambda -  g_{\mu\nu} g^{\rho \lambda} ) \bigg[ D_\rho \ti \phi D_\lambda \phi  +   \frac{3}{2} r V_\rho V_\lambda  \ti \phi \phi \nn\\
& &   \qquad \qquad  \qquad\qquad  \left.     +    (V_\rho + \kappa \left(\epsilon-1 \right)   K_\rho \right)\left(  \ii D_\lambda \ti \phi \, \phi  - \ii \ti \phi D_\lambda \phi  \right)   \bigg]\nn\\
 &  & + \frac{r}{2}  \left(    R_{\mu\nu} -   \frac12   g_{\mu\nu} R \right) \ti \phi \phi +  \frac{r}{2} \Big[  g_{ \mu \nu } \nabla_{\rho}\nabla^{\rho}  \big(  \ti \phi \phi   \big)     -   \nabla_{\mu} \nabla_{\nu} \big( \ti \phi \phi \big) \Big] \nn\\
 &  & +\frac{\ii}{2}   D_{(\mu}  \ti \psi \ti \sigma_{\nu)}  \psi - \frac{\ii}{2} \ti \psi  \ti \sigma_{(\mu}  D_{\nu)} \psi   -  \left( \frac12 V_{(\mu} + \kappa (1-\epsilon) K_{(\mu}  \right)\ti \psi  \ti \sigma_{\nu)}  \psi    ~ ,  \label{EM}
\eea
where the lower parenthesis denote symmetrization of the indices.  Recall that we defined $D_\mu =\nabla_\mu - \ii q_R A_\mu$, with $q_R$ the $R$ charges of the fields   \cite{Assel:2014paa}.

Below we collect some useful formulas for deriving this expression. 
For the bosonic part we used the variation of the Ricci tensor, 
\be
g^{\mu\nu} \delta {R}_{\mu \nu} \ = \  g_{ \mu \nu }  \nabla^{\rho} \nabla_{\rho} (\delta g^{ \mu \nu } ) -    \nabla_{\mu}\nabla_{\nu} (\delta g^{ \mu \nu} )  ,
\ee
and we note that for any vector field $X^\mu$, 
\be
\left[ \nabla_\mu , \nabla_\nu \right] X^\mu \ = \ R_{\mu\nu}   X^\mu ~.
\ee 
For the femionic part, the variation of the action with respect to the metric gives
\bea
\delta S^\mathrm{chiral}_\mathrm{fer} & = & \int \dd^4x \left[ \delta \sqrt{-g} \, \mathcal L^\mathrm{chiral}_\mathrm{fer} +  \sqrt{-g} \, \delta \mathcal L^\mathrm{chiral}_\mathrm{fer} \right]  \ = \ \int \dd^4x \sqrt{-g} \   \delta \mathcal L^\mathrm{chiral}_\mathrm{fer}    ~ , \label{variationofaction}
\eea
where in the second equality we used that $\mathcal L^\mathrm{chiral}_\mathrm{fer}$ vanishes on-shell. The variation of the Lagrangian can be expressed in terms of variations of the vielbein and of the spin connection and reads 
\bea
\delta \mathcal L^\mathrm{chiral}_\mathrm{fer} & = &  \ti \psi \ti \sigma^a  \left( \ii D_\mu + \frac12 V_\mu + \kappa(1-\epsilon) K_\mu \right) \psi \, \delta {e_a}^\mu -\frac{\ii}{2} \ti \psi \ti \sigma^\mu   \sigma^{ab} \psi \, \delta \omega_{\mu a b} ~ . \label{variationFerm}
\eea

Using the property  that the vielbein is covariantly constant,
\be
0 \ = \  \nabla_\mu {e_\nu}^a \ = \  \de_\mu  {e_\nu}^a - \Gamma_{\mu \nu}^\rho {e_\rho}^a + {{\omega_\mu}^a}_b {e_\nu}^b  ~ ,
\ee
we read off the variation of the spin connection
\be
\delta \omega_{\mu ab}  \ = \  \delta \Gamma_{\mu \nu}^\rho e_{a\rho} {e_b}^\nu - {e_b}^\nu \nabla_\mu ( \delta e_{a \nu} ) ~ .
\ee
Further, using the variation of the Christoffel symbol 
\be
\delta \Gamma^\sigma_{\mu\nu} \ = \   \frac12 g_{\mu \lambda} g_{\nu \rho} \nabla^\sigma (\delta g^{\lambda \rho} ) - g_{\lambda (\mu } \nabla_{\nu)} (\delta g^{\lambda \sigma} ) ~ ,
\ee
and
\be
\delta e_{a \nu} \ = \ -  g_{\nu \beta}{e_{a\alpha}} \delta g^{\alpha\beta} + g_{\mu\nu} \delta {e_a}^\mu ~ ,
\ee
we can write the variation of the spin connection as 
\be
\delta \omega_{\mu ab} \ = \  \nabla_\nu \left(  g_{\mu \lambda} {e_{[a}}^{\nu} e_{b] \rho}  \delta g^{\lambda \rho} + \frac12 {e_{a\lambda}} {e_{b\rho}}   \delta^\nu_\mu  \delta g^{\lambda\rho} - {e_{b \rho}} \delta_\mu^\nu  \delta {e_a}^\rho   \right)  ~  .
\ee
Using this, the second term of \eqref{variationFerm} can be written as,
\bea
-\frac{\ii}{2} \ti \psi \ti \sigma^\mu  \sigma^{ab} \psi  \, \delta \omega_{\mu a b}& =&  -  \ti \psi \ti \sigma^{a}   \left( \ii D_\mu + \frac12 V_\mu + \kappa(1-\epsilon) K_\mu \right)  \psi \,    \delta {e_a}^\mu  \nn\\
&& + \frac{\ii}{4} \left[   D_{\mu} \ti \psi  \ti \sigma_{\nu}\psi  -  \ti \psi  \ti \sigma_{\mu}     D_{\nu} \psi  \right] \delta g^{\mu\nu} \nn\\
&& - \frac12 \   \left( \frac12 V_{\mu} + \kappa (1-\epsilon) K_{\mu} \right)  \ti \psi  \ti \sigma_{\nu}  \psi \,  \delta g^{\mu\nu}  ~ ,
\eea
up to a total divergence. Substituting this back into \eqref{variationFerm}, the terms containing $\delta {e_a}^\mu$ cancel. The remaining terms are all proportional to $\delta g^{\mu\nu}$ and give the fermionic part of the energy-momentum tensor \eqref{EM}.

We also used the following identities for the $\sigma$-matrices in Lorentzian signature
\bea
\sigma^a \ti \sigma^b \sigma^c & = & - \eta ^{ab} \sigma^c + \eta^{ac} \sigma^b - \eta ^{bc} \sigma^a + \ii \epsilon^{abcd} \sigma_{d} ~,   \nn\\
\ti \sigma^a  \sigma^b \ti \sigma^c & = & - \eta ^{ab} \ti \sigma^c + \eta^{ac} \ti \sigma^b - \eta ^{bc} \ti \sigma^a - \ii \epsilon^{abcd} \ti \sigma_{d} ~, 
\eea
with $\epsilon^{0123} = -1$, and the identities
\bea
\left[ \nabla_\mu , \nabla_\nu \right]  \psi  & = & - \frac12 R_{\mu \nu ab}  \sigma^{a b}  \psi ~ , \qquad  \left[ \nabla_\mu , \nabla_\nu \right] \ti \psi \ = \ - \frac12 R_{\mu \nu ab}  \ti \sigma^{a b} \ti \psi  ~ ,
\eea
valid for generic spinors $\psi$, $\ti\psi$.

One can easily compute the  Ferrara-Zumino current
\be
\JFZ^\mu  \ = \  - \frac{2}{3} \frac{1}{\sqrt{-g}} \frac{\delta S}{\delta V_\mu} ~ ,
\ee
and the current
\be
J_K^\mu = \frac{1}{\sqrt{-g }} \frac{\delta S}{\delta K_\mu} ~ .
\ee
These read
\bea
 J_\mathrm{FZ}^\mu &=&  - \frac23 \left( \ii \ti \phi D^\mu  \phi   -  \ii D^\mu \ti \phi \, \phi   -3r V^\mu  \ti \phi \phi +\frac12 \ti \psi \ti \sigma^\mu \psi  \right) ~ , \\
 J_K^\mu &=&  \kappa(1-\epsilon)  \Big(   \ii D^\mu \ti \phi \, \phi  -  \ii   \ti \phi D^\mu  \phi  + \ti \psi \ti \sigma^\mu \psi  \Big) ~ ,
\eea
and are not conserved. Starting with the expressions above, a further computation yields (\ref{Tnoncons}).


\begin{thebibliography}{9}

\bibitem{Witten:1988ze} 
  E.~Witten,
  \emph{Topological Quantum Field Theory},
  Commun.\ Math.\ Phys.\  {\bf 117}, 353 (1988).
 
  
\bibitem{Assel:2014paa} 
  B.~Assel, D.~Cassani and D.~Martelli,
  \emph{Localization on Hopf surfaces},
  JHEP {\bf 1408}, 123 (2014)
  [arXiv:1405.5144 [hep-th]].

 
\bibitem{Closset:2013vra} 
  C.~Closset, T.~T.~Dumitrescu, G.~Festuccia and Z.~Komargodski,
  \emph{The Geometry of Supersymmetric Partition Functions},
  JHEP {\bf 1401}, 124 (2014)
  [arXiv:1309.5876 [hep-th]].


\bibitem{Romelsberger:2005eg} 
  C.~Romelsberger,
  \emph{Counting chiral primaries in ${\cal N} = 1$, $d=4$ superconformal field theories},
  Nucl.\ Phys.\ B {\bf 747}, 329 (2006)
  [hep-th/0510060].



\bibitem{Cassani:2014zwa} 
  D.~Cassani and D.~Martelli,
  \emph{The gravity dual of supersymmetric gauge theories on a squashed $S^1\times S^3$},
  JHEP {\bf 1408}, 044 (2014)
  [arXiv:1402.2278 [hep-th]].


\bibitem{Bloete:1986qm} 
  H.~W.~J.~Bloete, J.~L.~Cardy and M.~P.~Nightingale,
  \emph{Conformal Invariance, the Central Charge, and Universal Finite Size Amplitudes at Criticality},
  Phys.\ Rev.\ Lett.\  {\bf 56}, 742 (1986).



\bibitem{Cappelli:1988vw} 
  A.~Cappelli and A.~Coste,
  \emph{On the Stress Tensor of Conformal Field Theories in Higher Dimensions},
  Nucl.\ Phys.\ B {\bf 314}, 707 (1989).



\bibitem{Herzog:2013ed} 
  C.~P.~Herzog and K.~W.~Huang,
  \emph{Stress Tensors from Trace Anomalies in Conformal Field Theories},
  Phys.\ Rev.\ D {\bf 87}, 081901 (2013)
  [arXiv:1301.5002 [hep-th]].



\bibitem{Brown:1977sj} 
  L.~S.~Brown and J.~P.~Cassidy,
  \emph{Stress Tensors and their Trace Anomalies in Conformally Flat Space-Times},
  Phys.\ Rev.\ D {\bf 16}, 1712 (1977).



\bibitem{Assel:2014tba} 
  B.~Assel, D.~Cassani and D.~Martelli,
  \emph{Supersymmetric counterterms from new minimal supergravity},
  JHEP {\bf 1411}, 135 (2014)
  [arXiv:1410.6487 [hep-th]].



\bibitem{Sohnius:1981tp} 
  M.~F.~Sohnius and P.~C.~West,
  \emph{An Alternative Minimal Off-Shell Version of ${\cal N}=1$ Supergravity},
  Phys.\ Lett.\ B {\bf 105}, 353 (1981).



\bibitem{Kim:2012ava} 
  H.~C.~Kim and S.~Kim,
  \emph{M5-branes from gauge theories on the 5-sphere},
  JHEP {\bf 1305}, 144 (2013)
  [arXiv:1206.6339 [hep-th]].



\bibitem{Buican:2014qla} 
  M.~Buican, T.~Nishinaka and C.~Papageorgakis,
  \emph{Constraints on chiral operators in $ \mathcal{N}=2 $ SCFTs},
  JHEP {\bf 1412}, 095 (2014)
  [arXiv:1407.2835 [hep-th]].



\bibitem{DiPietro:2014bca} 
  L.~Di Pietro and Z.~Komargodski,
  \emph{Cardy Formulae for SUSY Theories in $d=4$ and $d=6$},
  JHEP {\bf 1412} 031 (2014),
  [arXiv:1407.6061 [hep-th]].



\bibitem{Ardehali:2014esa} 
  A.~A.~Ardehali, J.~T.~Liu and P.~Szepietowski,
  \emph{Central charges from the $\mathcal{N} =$ 1 superconformal index},
  Phys.\ Rev.\ Lett.\  {\bf 114}, no. 9, 091603 (2015)
  [arXiv:1411.5028 [hep-th]].


\bibitem{birrell}
N.~ D.~ Birrell and P.~C.~W.~ Davies,
\emph{Quantum fields in curved space},
Cambridge University Press (1982).


\bibitem{Marino:2011nm} 
  M.~Marino,
  \emph{Lectures on localization and matrix models in supersymmetric Chern-Simons-matter theories},
  J.\ Phys.\ A {\bf 44}, 463001 (2011)
  [arXiv:1104.0783 [hep-th]].



\bibitem{Assel:2015nca} 
  B.~Assel, D.~Cassani, L.~Di Pietro, Z.~Komargodski, J.~Lorenzen and D.~Martelli,
  \emph{The Casimir Energy in Curved Space and its Supersymmetric Counterpart},
  [arXiv:1503.05537 [hep-th]].




\bibitem{Sen:1985dc} 
  D.~Sen,
  \emph{Fermions in the space-time $\mathbb{R} \times S^3$},
  J.\ Math.\ Phys.\  {\bf 27}, 472 (1986).



\bibitem{Romelsberger:2007ec} 
  C.~Romelsberger,
  \emph{Calculating the Superconformal Index and Seiberg Duality},
  [arXiv:0707.3702 [hep-th]].

  

\bibitem{Closset:2013sxa} 
  C.~Closset and I.~Shamir,
  \emph{The $\mathcal{N}=1$ Chiral Multiplet on $T^2\times S^2$ and Supersymmetric Localization},
  JHEP {\bf 1403}, 040 (2014)
  [arXiv:1311.2430 [hep-th]].

       
\bibitem{Kapustin:2009kz} 
  A.~Kapustin, B.~Willett and I.~Yaakov,
  \emph{Exact Results for Wilson Loops in Superconformal Chern-Simons Theories with Matter},
  JHEP {\bf 1003}, 089 (2010)
  [arXiv:0909.4559 [hep-th]].

  
\bibitem{Jafferis:2010un} 
  D.~L.~Jafferis,
  \emph{The Exact Superconformal $R$-Symmetry Extremizes $Z$},
  JHEP {\bf 1205}, 159 (2012)
  [arXiv:1012.3210 [hep-th]].



\bibitem{Hama:2010av} 
  N.~Hama, K.~Hosomichi and S.~Lee,
  \emph{Notes on SUSY Gauge Theories on Three-Sphere},
  JHEP {\bf 1103}, 127 (2011)
  [arXiv:1012.3512 [hep-th]].



\bibitem{Festuccia:2011ws} 
  G.~Festuccia and N.~Seiberg,
  \emph{Rigid Supersymmetric Theories in Curved Superspace},
  JHEP {\bf 1106}, 114 (2011)
  [arXiv:1105.0689 [hep-th]].


\bibitem{Elizalde:2012zza}
    E. Elizalde, \emph{Ten physical applications of spectral zeta functions}, Lect.  Notes  Phys. \textbf{855}, Springer  (2012). 
       

\bibitem{Dowker:1998tb} 
  J.~S.~Dowker,
  \emph{On the relevance of the multiplicative anomaly},
  [hep-th/9803200].

  


\bibitem{McKenzieSmith:2000vz} 
  J.~J.~McKenzie-Smith and D.~J.~Toms,
  \emph{Zero point energies and the multiplicative anomaly},
  [hep-th/0005201].




\bibitem{Page:1982fm} 
  D.~N.~Page,
  \emph{Thermal Stress Tensors in Static Einstein Spaces},
  Phys.\ Rev.\ D {\bf 25}, 1499 (1982).



\bibitem{Cutkosky:1983jd} 
  R.~E.~Cutkosky,
  \emph{Harmonic Functions and Matrix Elements for Hyperspherical Quantum Field Models},
  J.\ Math.\ Phys.\  {\bf 25}, 939 (1984).


\end{thebibliography}
\end{document}